\documentclass[journal,comsoc]{IEEEtran}

\usepackage{url}
\usepackage{adjustbox}
\usepackage{multirow}
\usepackage{array}
\usepackage{graphicx}
\usepackage{booktabs} 

\usepackage{cite}

\ifCLASSINFOpdf
\else
\fi
\begin{document}
%
\title{Interoperability of the Metaverse: \\A Digital Ecosystem Perspective Review}
%
%
%

\author{Liang~Yang,
        Shi-Ting~Ni,
        Yuyang~Wang,~\IEEEmembership{Member,~IEEE,}
        Ao~Yu,
        Jyn-An~Lee,
    and~Pan~Hui,~\IEEEmembership{Fellow,~IEEE}
\thanks{L. Yang is with Division of Emerging Interdisciplinary Areas, The Hong Kong University of Science and Technology; and Computational Media and Arts Thrust, The Hong Kong University of Science and Technology (Guangzhou), e-mail: lyangbl@connect.ust.hk.}
\thanks{S. Ni, Y. Wang and A. Yu are with Computational Media and Arts Thrust, The Hong Kong University of Science and Technology (Guangzhou).}
\thanks{J. Lee is with Centre for Legal Innovation and Digital Society(CLINDS), The Chinese University of Hong Kong Faculty of Law.}
\thanks{P. Hui is with Computational Media and Arts Thrust, The Hong Kong University of Science and Technology (Guangzhou); Division of Emerging Interdisciplinary Areas, The Hong Kong University of Science and Technology; Department of Computer Science, University of Helsinki.}}
\maketitle

\begin{abstract}
The Metaverse is at the forefront of the digital revolution, with significant potential to transform industries and lifestyles. However, skepticism has emerged within industrial and academic circles, raising concerns that enthusiasm may outpace technological advancements. Interoperability has been identified as a major barrier to realizing its full potential. A report by CoinMarketCap noted that, as of February 2023, over 240 Metaverse initiatives existed in isolation, highlighting this challenge. Despite consensus on the importance of interoperability, systematic research on this theme is lacking. Our study addresses this gap through a systematic literature review, employing content analysis methodologies in a structured search of the Web of Science and Scopus databases, yielding 74 pertinent publications. Interoperability is challenging to define due to varying contexts and lack of a standardized definition. Similarly, the Metaverse lacks a uniform definition but is generally viewed as a digital ecosystem. Urs Gasser proposed a framework for interoperability within digital ecosystems, outlining technological, data, human, and institutional dimensions. Incorporating Gasser's framework, we analyze the literature within our identified three layers to offer a comprehensive overview of Metaverse interoperability research. Our study aims to set benchmarks for future inquiries, guiding this complex field and contributing to its scholarly development.
\end{abstract}


\section{Introduction}

\begin{figure*}
    \centering
    \includegraphics[width=\textwidth]{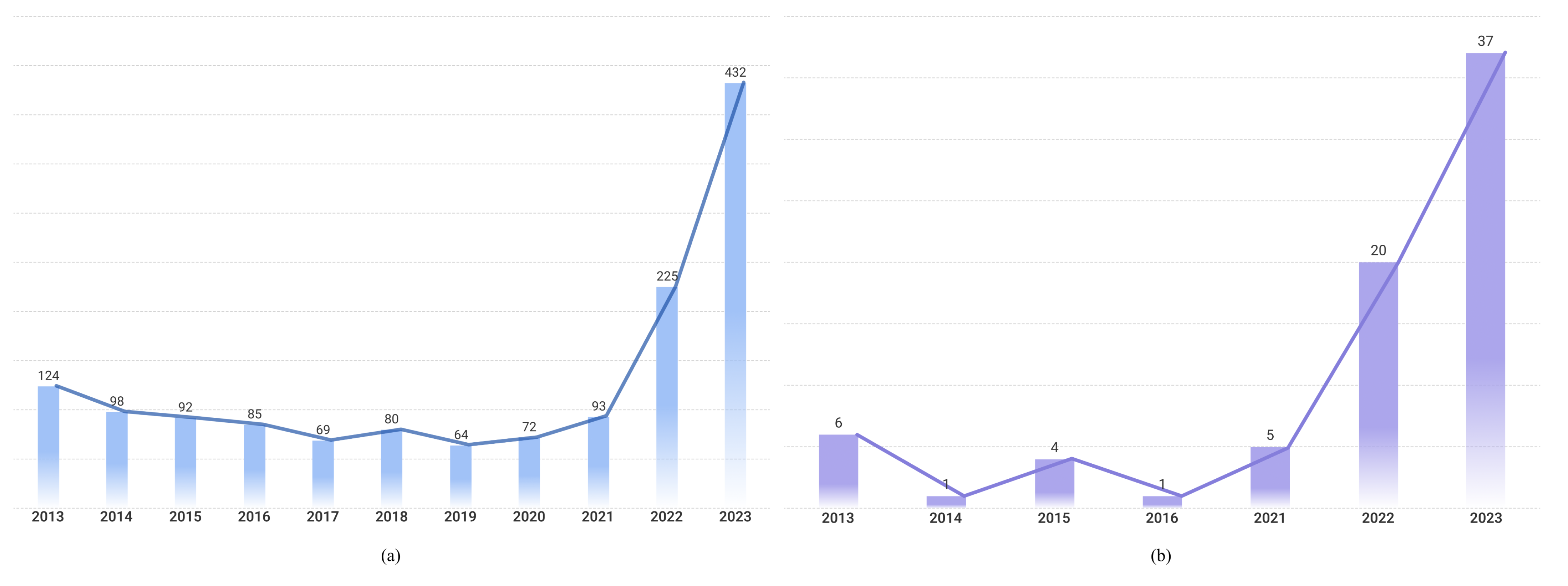}
    \caption{(a) Metaverse Research Trends 2012-2023: English Articles and Proceedings from WoS and Scopus 
    (b) Metaverse Interoperability Literature Histogram by Year}
    \label{fig:Figure.1}
\end{figure*}

The concept of the Metaverse has rapidly evolved from speculative fiction into a tangible goal for future digital interaction. Envisioned as a collective virtual shared space merging enhanced physical and digital realities, the Metaverse promises unprecedented  opportunities for social engagement, entertainment, and commerce~\cite{saritas2022systematic}. An analysis of publication trends over the past decades, depicted in Figure~\ref{fig:Figure.1}(a), reveals a steep upward trajectory, indicating rapid growth in Metaverse research. This surge underscores the increasing recognition of the Metaverse as a critical area of inquiry and innovation. 

Despite its widespread discussion, the term "Metaverse" lacks a specific, universal definition, as revealed by a detailed literature review~\cite{almoqbel2022metaverse}. Dionisio et al. describe it as a realistic, immersive environment that facilitates omnipresent access, consistent identity, interoperability, and scalability~\cite{dionisio20133d}. The Metaverse is often referred to as the next evolution of the Internet, typically described as a network of interconnected, immersive digital spaces accessible through multiple devices~\cite{weinberger2022metaverse}. Matthew Ball defines the Metaverse as a vast, interoperable network of real-time 3D virtual worlds where countless users maintain personal presence and consistent data, such as identity, history, and assets~\cite{ball2022metaverse}. Lee et al. conceptualize the Metaverse as a digitally created space merging aspects of the physical and virtual worlds~\cite{Lee2021AllON}. Abilkaiyrkyzy et al. describe the Metaverse as a universe composed of persistent digital twins—virtual counterparts of both living and non-living entities~\cite{Akbobek2023MetaPlatform}. A common emphasis in mainstream definitions is on interoperability—the seamless interconnection capability. However, these interoperability challenges also constitute a significant impediment to the Metaverse's advancement~\cite{wang2022survey, ball2022metaverse,shakila2023layerwiseI}.

In 2023, leaders from industrial and academic sectors emphasized the significance of these challenges. The World Economic Forum and Accenture highlighted the necessity of interoperability for transforming social interactions and the digital economy in their white paper "Interoperability in the Metaverse," advocating for robust governance and collaborative efforts~\cite{WEF_Intero_Metaverse}. Harvard Business School scholars Andy Wu and David B. Yoffie underscored a critical debate: will the Metaverse evolve as an open, interconnected ecosystem akin to the Internet, or will it resemble a collection of siloed platforms similar to today's app stores and social networks? The resolution of this debate is not merely theoretical; it carries substantial economic consequences, with the potential to direct billions in investment and shape the financial landscapes of the future~\cite{wu2023metaverse}.

Despite the launch of over 240 metaverse initiatives by February 2023, as reported by CoinMarketCap, the majority remain standalone endeavors, underscoring the urgent necessity for interoperability~\cite{jiang2023fast}. In recent years, there has been a notable surge in academic literature discussing Metaverse interoperability, as illustrated by Figure \ref{fig:Figure.1} (b). However, despite the growing awareness, there remains a discernible gap in scholarly literature: a comprehensive assessment of the complex nature of interoperability research within the Metaverse, including its definitions and scope, current status, possible solutions, and the broader pathway forward. This study aims to bridge this gap by proposing four research questions (RQs) designed to unravel the intricacies of achieving a fully interoperable Metaverse:

\begin{itemize}
  \item \textbf{RQ1:} How should the analysis of Metaverse interoperability be framed?
  \item \textbf{RQ2:} What consensus themes exist within the current literature on Metaverse interoperability? 
  \item \textbf{RQ3:} What are the main research findings within these themes, and how can they be systematically integrated?
  \item \textbf{RQ4:} What are the current challenges and future research agendas?

\end{itemize} 

This study employs a methodical approach to address our research questions by analyzing 74 articles spanning two decades (2003-2023). Using content analysis methodologies, we aim to discern patterns, identify gaps, decode discussions on this topic, and understand the evolving narrative of the Metaverse interoperability landscape. The investigation is divided into specific sections addressing the research questions. Section 2 sets the scene by providing an overview of interoperability in the digital ecosystem and its significance in the Metaverse context, addressing RQ1. Section 3 outlines our comprehensive review strategy and research methodology, reporting interim results. Section 4 presents a detailed literature analysis, categorizing key themes and integrating research outcomes to answer RQ2 and RQ3. Section 5 synthesizes these insights to highlight current obstacles and propose an agenda for future research, addressing RQ4. The paper concludes with Section 6, summarizing our findings and implications for the field.

\begin{figure*}
    \centering
    \includegraphics[width=\textwidth]{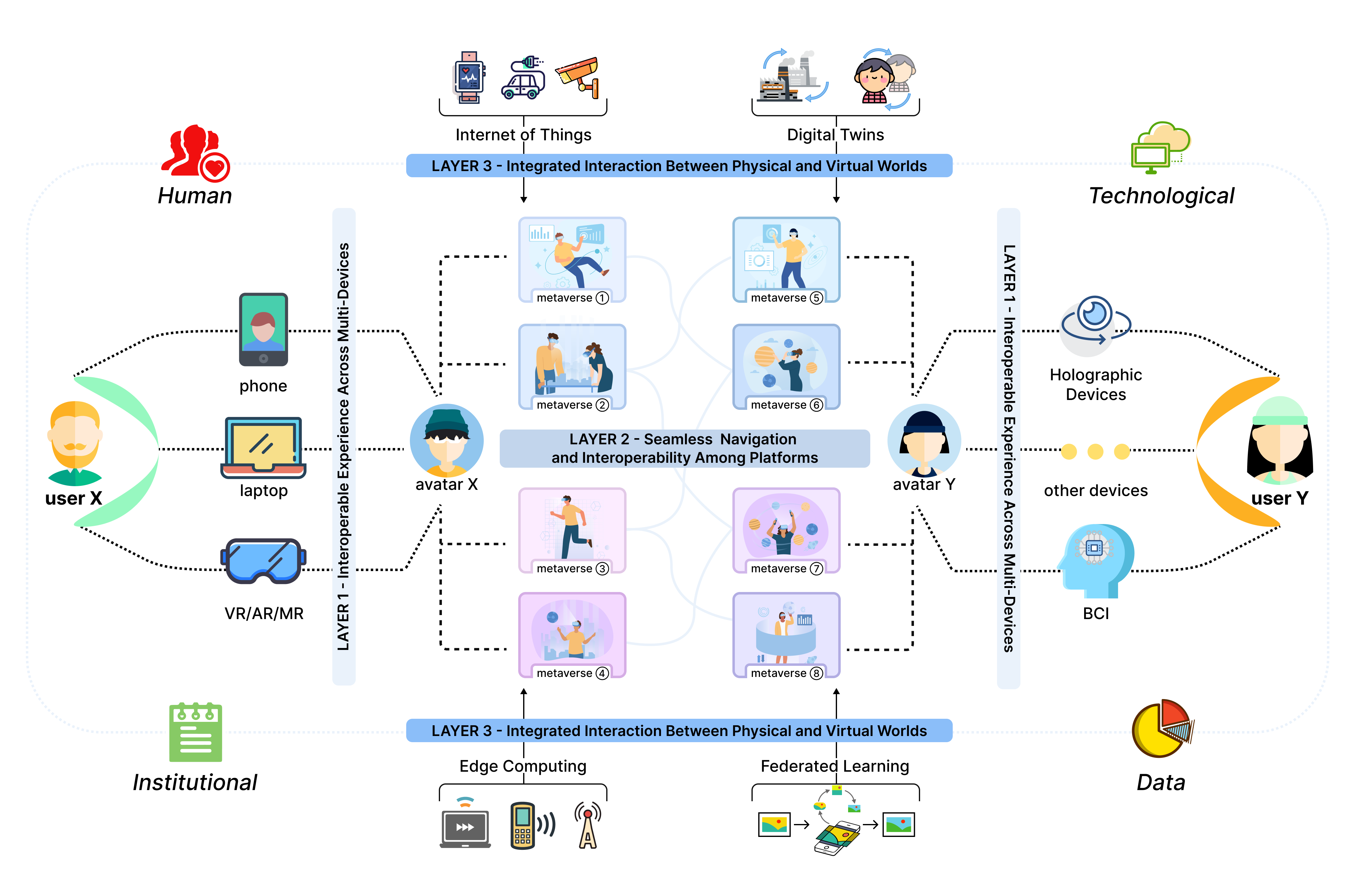}
    \caption{The Scope Visualization of Metaverse Interoperability Based on Our Research Findings}
    \label{fig: full map}
\end{figure*}

Our contributions are as follows: Firstly, we have identified a critical gap in the literature—a lack of comprehensive, systematic reviews on the topic of Metaverse interoperability, despite its growing attention. Our work pioneers in bridging this gap. Secondly, as the Metaverse emerges as a new-generation digital ecosystem, and interoperability presents a multi-faceted and broad topic, we have devised a theoretical framework to organize and deepen our understanding of this complexity. Furthermore, our literature review has pinpointed four dimensions within three key layers for different aspects of this subject. We offer an in-depth analysis of these dimensions and layers, providing fresh insights and a structured approach to comprehend the full scope of the Metaverse interoperability, aiding in achieving greater consensus on its concepts and extent within both academic and industrial spheres, as depicted in Figure \ref{fig: full map}. We clearly outline the present challenges and establish a research agenda, setting the stage for continuous academic exploration and technological innovation in the realm of Metaverse interoperability.


\section{Background}
This section provides an overview of the scope and definition of Metaverse interoperability. It then introduces Urs Gasser's interoperability framework, structured across four layers: technological, data, human, and institutional\cite{gasser2015intero}. This framework is pertinent to our analysis as it offers a comprehensive lens to examine the multifaceted nature of interoperability within the Metaverse. We conclude by detailing how we apply this framework to the current research.

\subsection{Interoperability in the Digital Ecosystem}
Interoperability is a complex and multifaceted concept~\cite{ouksel1999semantic,hatzi2018intero}, encompassing technical, semantic, organizational, and legal dimensions~\cite{WBID4D}. From a technical perspective, ISO/IEC/IEEE 24765:2017 defines interoperability as the ability of systems, products, or components to exchange information effectively and to use that information~\cite{ISO24765}. This includes seamless cooperation between object request brokers, data transfer with minimal user knowledge of system intricacies, and the collaborative capability of objects to communicate and exchange actionable information. The ISO/IEC/IEEE standards emphasize the significance of technical interoperability for smooth information flow and collaborative functionality. 
Beyond technical implementation, semantic interoperability is crucial for ensuring that shared information is accurately understood and interpreted across different systems~\cite{ouksel1999semantic, heiler1995semantic}. It involves aligning the meaning and interpretation of data, messages, and commands among systems and interpreting knowledge from other languages at the semantic level, assigning the correct interpretation or set of models to each piece of imported knowledge~\cite{Euzenat2001TowardsAP}. Achieving semantic interoperability is essential to prevent misunderstandings, enhance collaborative efficiency, and facilitate seamless integration across diverse systems.
Organizational and governance considerations are also integral to achieving interoperability. Organizations operate with unique business processes, data formats, and standards, which require establishing common criteria, protocols, and rules for cross-organizational interoperability~\cite{WBID4D}. Furthermore, appropriate governance mechanisms and collaborative frameworks are necessary to foster stakeholder cooperation and coordination and to address potential conflicts and challenges.
Interoperability is also influenced by legal and regulatory constraints, particularly concerning data protection, privacy, and security~\cite{Haugum2022SecurityAP, Correia2023SecurityAP}. Market competition and platform governance issues are relevant as well~\cite{morton2023equitable}. Achieving interoperability involves not only adhering to laws and regulations but also ensuring the compatibility of legal authorizations across various organizations for compliance and functional cooperation.

\begin{figure}
    \centering
    \includegraphics[width=1\linewidth]{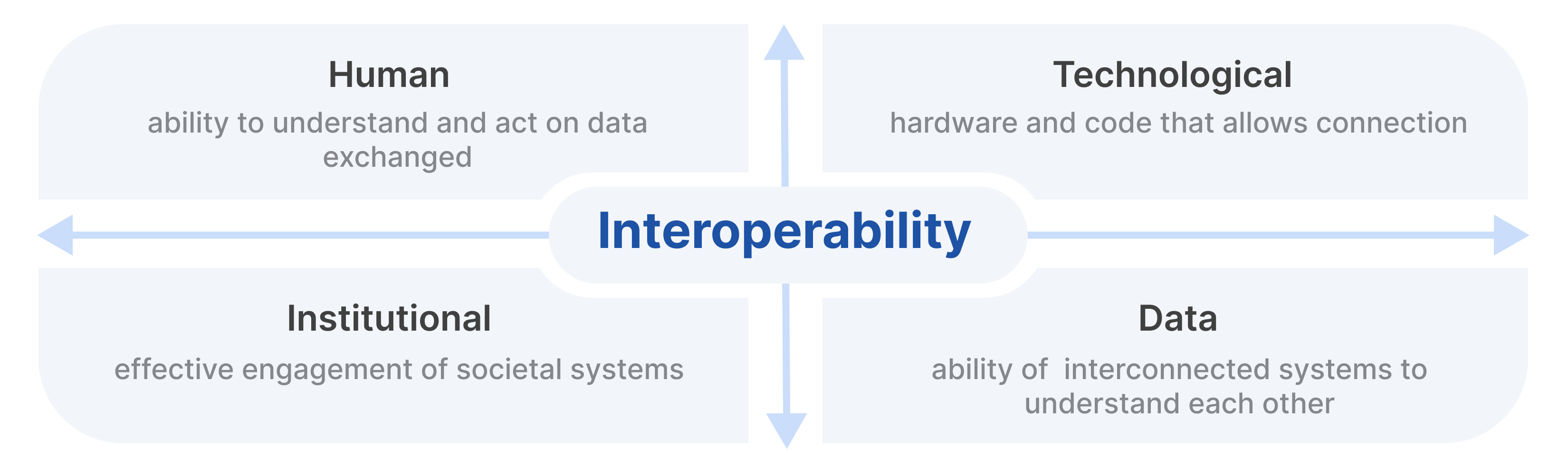}
    \caption{Urs Gasser's Framework for Understanding Interoperability within Digital Ecosystem}
    \label{fig:Gasser's Framework}
\end{figure}

The definition of interoperability is challenging due to its context-dependent nature and the lack of a universal definition. Fundamentally, within the realm of information technology, it represents the ability to exchange and utilize data and information across different systems, applications, or components. Urs Gasser provides a multifaceted framework for understanding interoperability, outlining technological, data, human, and institutional functional dimensions, as well as the associated benefits and potential risks~\cite{gasser2015intero}. This model offers structured approaches to navigate the complexities of interoperability, as illustrated in Figure \ref{fig:Gasser's Framework}. The technological dimension involves the hardware and code that enable physical connections between systems, facilitating data sharing via agreed-upon interfaces. The data dimension focuses on the ability of interconnected systems to understand and interpret shared data, closely intertwined with the technological dimension. The human dimension emphasizes the assurance of user experience in comprehending and effectively utilizing exchanged data. The institutional dimension extends into the realm of societal systems, encapsulating standards organizations, collaborative methods, as well as legal and policy frameworks, which do not require complete uniformity but rather a sufficient level of commonality to safeguard the interests of involved parties. Urs Gasser's framework provides a comprehensive approach to navigating the complexities of interoperability within a digital ecosystem. It is important to recognize that the division into these four dimensions merely establishes a conceptual framework; the categories are not entirely exclusive, and some unavoidable overlaps exist, particularly among the human, technical, and data dimensions, a nuance acknowledged in our application of this framework but does not detract from its overall utility.

\subsection{Interoperability in the Metaverse}
The Metaverse, which has yet to be uniformly defined, is widely recognized as the digital ecosystem that succeeds the Internet~\cite{ball2022metaverse, Cheng2022WillMB, ritterbusch2023defining, duong2023digital}. Our literature review categorizes discussions of the Metaverse into three primary layers. The first layer envisions the Metaverse as an evolved user experience driven by cutting-edge technologies such as Virtual Reality (VR), Augmented Reality (AR), and Mixed Reality (MR)~\cite{Mystakidis2019Metaverse, Lee2021AllON, ball2022metaverse, tumler2022multi}. These innovations facilitate a shift to spatial engagement, transitioning from 2D to 3D experiences. Rather than replacing existing digital platforms, they enhance them by merging with the current Internet infrastructure to provide a richer, more immersive form of interaction, predominantly through advanced smart devices~\cite{garcia2021cross, tumler2022multi, yunsik2023xave}.
The second layer forges a pervasive, immersive digital landscape that has evolved from the Internet and Massively Multiplayer Online Games (MMOGs)~\cite{lee2023legal, Soni2023ARO}. In this virtual realm, users navigate as avatars, engaging with various virtual content and environments that reflect the physical world. It offers a space for enhanced sensory engagement, self-expression, and presence~\cite{Rajan2002ARV,Schwartz2018TheIV}. This layer builds on advancements in VR, AR, and MR and also integrates Artificial Intelligence (AI)~\cite{ali2023federatedAI} and blockchain technology~\cite{huynh2023blockchain, Huang2023EconomicSys}, adding layers of security and trust to users' virtual lives and economic interactions.
The Metaverse envisions a seamless fusion of the physical and virtual worlds in the third layer, effectively dissolving the divide between digital and physical realities~\cite{yu2022ParallelSensing, wook2023standardization}. This advanced stage imagines life within a "surreality", a blended space where everyday living, work, and play are redefined across integrated realities. This integration extends beyond visual experiences to include tactile~\cite{pham2022tactile} and auditory sensations~\cite{jean2021rspatial}. Leveraging the Internet of Things (IoT)~\cite{Jie2022ERIoT}, digital twins~\cite{tu2023twinxr}, external or wearable devices~\cite{lal2021GiveH, Saeed2023SemCom, ali2023federatedAI}, and biometric technologies~\cite{wang2023survey}, the Metaverse aims to provide real-time feedback on users' physiological and emotional states, enabling highly personalized and responsive experiences~\cite{Jie2022EXRI}. This exploration promises to redefine human lifestyles and work paradigms.


From a functional standpoint, interoperability in the Metaverse resembles that within a universal digital ecosystem, as discussed in section 2.1. It involves the seamless and ideally transparent exchange of information and interactions between diverse systems or platforms, supported by a consensus that matures into formalized standards~\cite{dionisio20133d,wook2023standardization,Anita20233DWebInter}. Within the Metaverse context, recognizing interoperability as a key characteristic is critical; it determines whether the Metaverse can emerge as a truly unified space, deserving of a capitalized `M'. Just as the capitalized `I' in the Internet represents a global network ecosystem, the Metaverse requires a foundational understanding and communication between diverse platforms and spaces~\cite{dionisio20133d, ball2022metaverse}. The goal is not only maintaining technical compatibility but also crafting a seamless and intuitive user experience as the information exchange that the Internet facilitates.

We aim to synthesize the relevant interoperability literature and utilize Urs Gasser's framework, along with the three-layer structure we identified, to perform an applied analysis. This endeavor is set to yield a comprehensive and detailed map of interoperability within the Metaverse. The structured analysis aims to provide clear, actionable insights for the interoperable evolution of the Metaverse.


\section{Research Method}
\subsection{Database and Search Query}
The methodology of systematic reviews critically relies on the choice of search systems to ensure objectivity and replicability, a principle that is especially pertinent for nascent research areas such as Metaverse interoperability. To broaden the scope of research discovery, our research utilizes Web of Science (WoS) and Scopus exclusively to identify pertinent studies on the topic.

\begin{figure*}
    \centering
    \includegraphics[width=\textwidth]{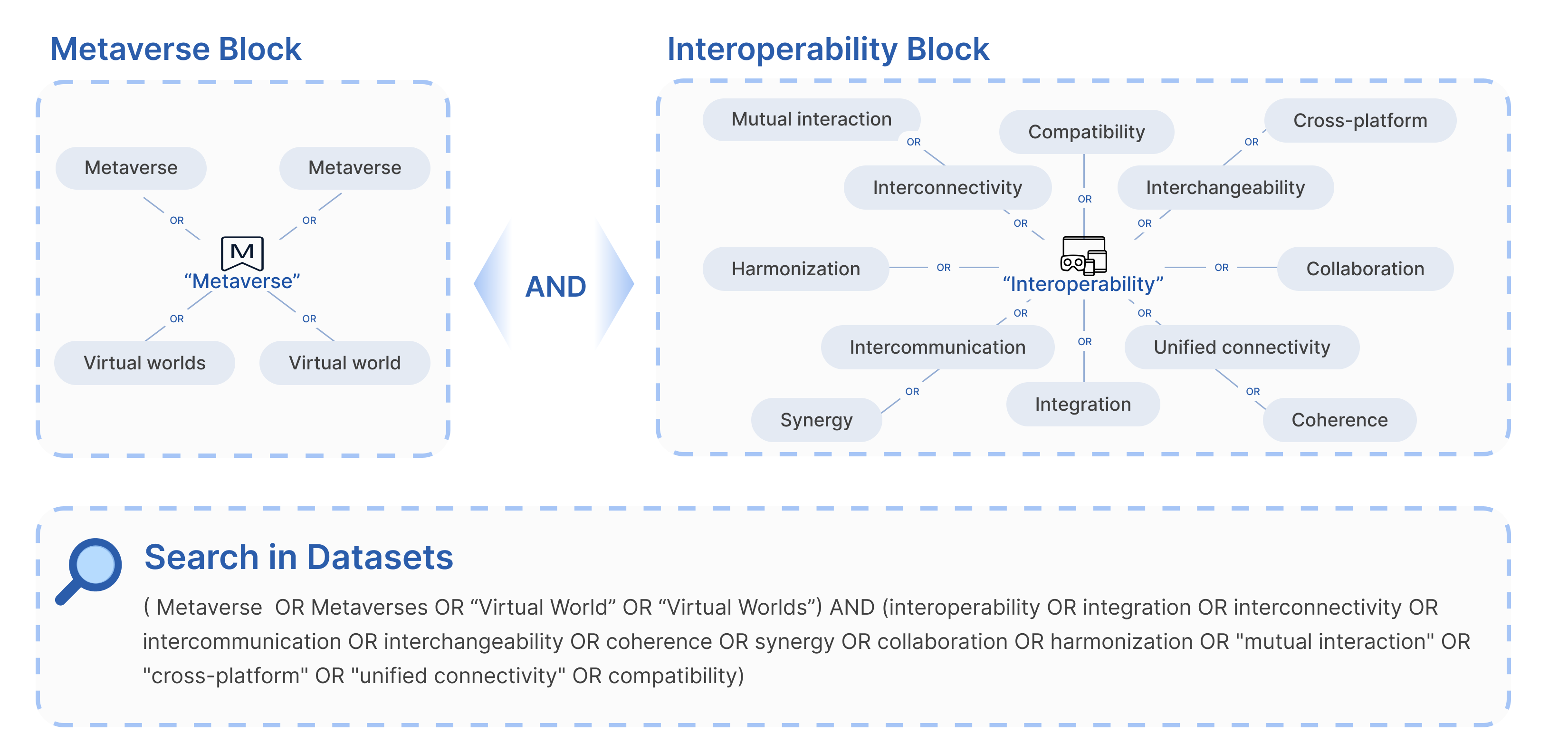}
    \caption{Process of Building the Search Query for Identifying Studies Related to Metaverse Interoperability }
    \label{fig:Process of Search Query}
\end{figure*}

The construction of a comprehensive search query is essential when embarking on a query-based search, particularly for a multifaceted concept like Metaverse. As depicted in Figure~\ref{fig:Process of Search Query}, our methodology for creating search queries to identify relevant literature on Metaverse interoperability involves two key terms: `Metaverse' and `Interoperability'. In line with the evolution of Metaverse discussed in Section 2.2, we include the precursor term `Virtual World' and connect it with `OR' to accommodate various stages of the Metaverse's development. Additionally, we expand the search for `Interoperability' by including synonyms such as `Interconnectivity' and `Intercommunication', linked with `OR'. The final search query is synthesized by combining these critical elements with `AND', thus ensuring a focused and comprehensive search strategy to uncover studies pertinent to interoperability within the Metaverse.

\subsection{Study Identification (Stage 1)}

\begin{figure*}
    \centering
    \includegraphics[width=\textwidth]{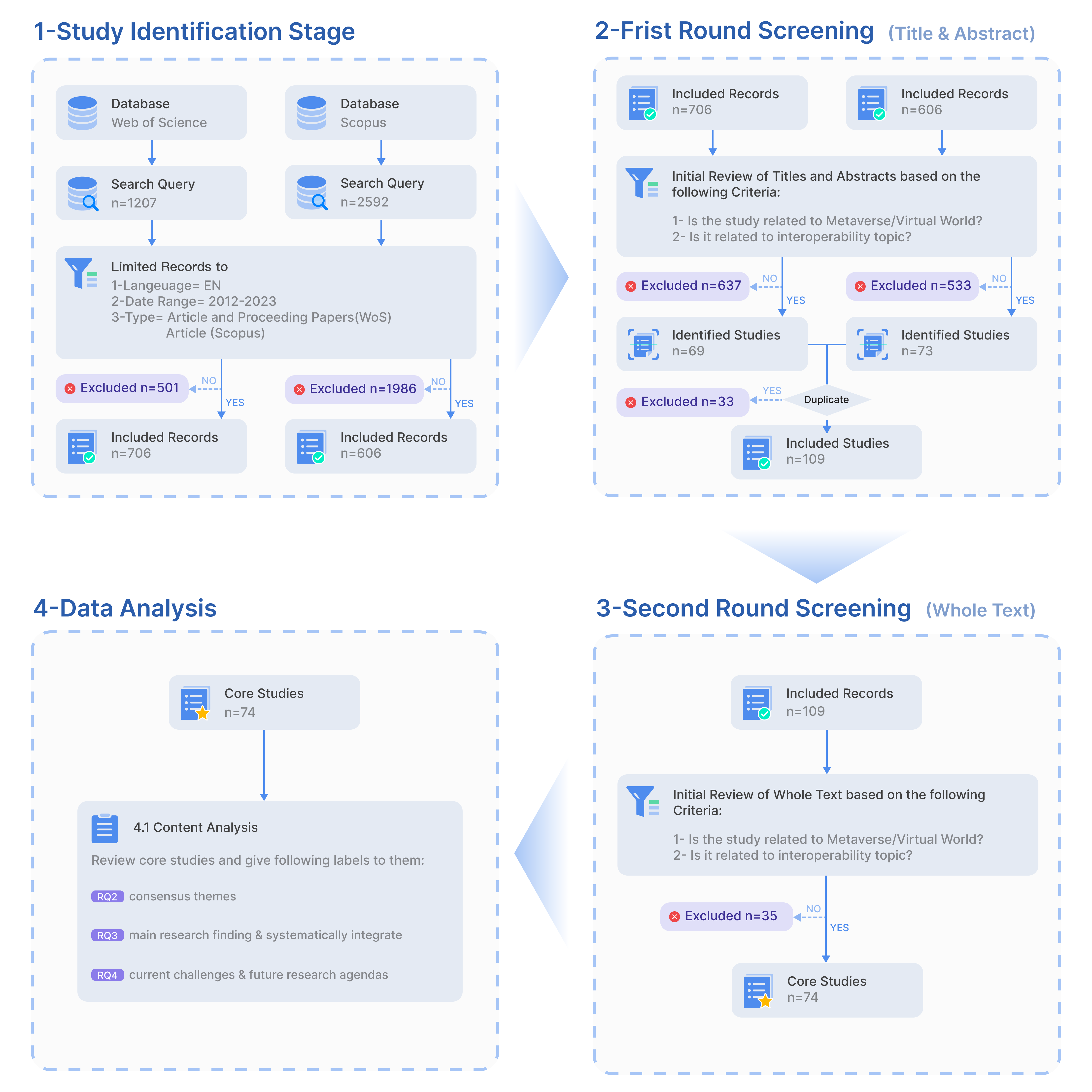}
    \caption{Stages of the Systematic Review of Metaverse Interoperability Literature}
    \label{fig:stages of review}
\end{figure*}

Upon establishing the groundwork for potential databases and constructing search queries, we commenced the identification of relevant studies. Figure~\ref{fig:stages of review} outlines the four stages of our review process, beginning with developing a search query. This query was applied to the "Title," "Abstract," and "Keywords" fields in the selected databases using the `OR' command, ensuring comprehensive retrieval of studies containing any of the query keywords within the WoS and Scopus databases. As of January 6, 2024, the number of records identified was 1,207 in WoS and 2,592 in Scopus.

We applied several filters to refine the search results and ensure relevance to our review's scope. Firstly, we limited the records to those available in English. Secondly, given Metaverse's relatively recent emergence, as discussed in the previous section, we restricted our search to the past decade, from 2013 to 2023. Since journal articles are considered highly influential in systematic reviews~\cite{santos2017bibliometric}, and given the importance of conference papers in disseminating computer science research—a field central to Metaverse technologies~\cite{vardi2009conferences}—we included both journal articles and conference proceedings in WoS. However, in Scopus, we limited the search to journal articles for two reasons: Scopus indexes fewer proceedings papers compared to WoS, and there is considerable overlap between the two databases~\cite{kokol2018discrepancies}. Additionally, conference papers often provide preliminary findings rather than substantially contributing to the body of knowledge. These criteria reduced the number of relevant records to 706 in WoS and 606 in Scopus.

\subsection{Screening (Stage 2 and 3)}
Continuing the review process, we established two key inclusion criteria to evaluate the relevance of studies on Metaverse interoperability: (1) the study must pertain to the Virtual Worlds and Metaverse sector, and (2) the study must address interoperability, whether through technological, organizational, or governance and policy dimensions. This specificity was required due to the broad and multifaceted nature of interoperability. As depicted in Figure~\ref{fig:stages of review}, the second stage involves screening study titles and abstracts obtained during the initial search against these criteria, leading to the exclusion of 637 studies from WoS and 533 from Scopus. Subsequently, 33 duplicate records were identified and removed, reducing the studies to 109. Where titles and abstracts did not provide enough information to assess relevance, full-text assessments were conducted, as shown in Stage 3 of Figure~\ref{fig:stages of review}. Ultimately, 74 studies fulfilled the inclusion criteria and were selected for detailed analysis.

\subsection{Data Analysis Methods (Stage 4)}
In Stage 4, we applied content analysis to investigate the research questions. We adopted Wook Hyun's classification of Metaverse interoperability~\cite{wook2023standardization}, distinguishing between vertical and horizontal layers. The horizontal layer involves interactions among different Metaverse platforms or with third-party service platforms, while the vertical layer relates to integration with the real world~\cite{wook2023standardization}. Expanding on our content analysis of the 74 key studies, we further developed Hyun's frameworks into three distinct layers to address RQ2. Employing Urs Gasser's theoretical framework, we recognized that interoperability encompasses \textit{human}, \textit{technical}, \textit{data}, and \textit{institutional}~\cite{gasser2015intero} dimensions. Our analysis indicated clear distinctions in these four dimensions within our three interoperability layers. We meticulously extracted and categorized the principal themes of interoperability from the core studies.



\section{Literature Analysis and Findings}

\begin{figure}
    \centering
    \includegraphics[width=0.8\linewidth]{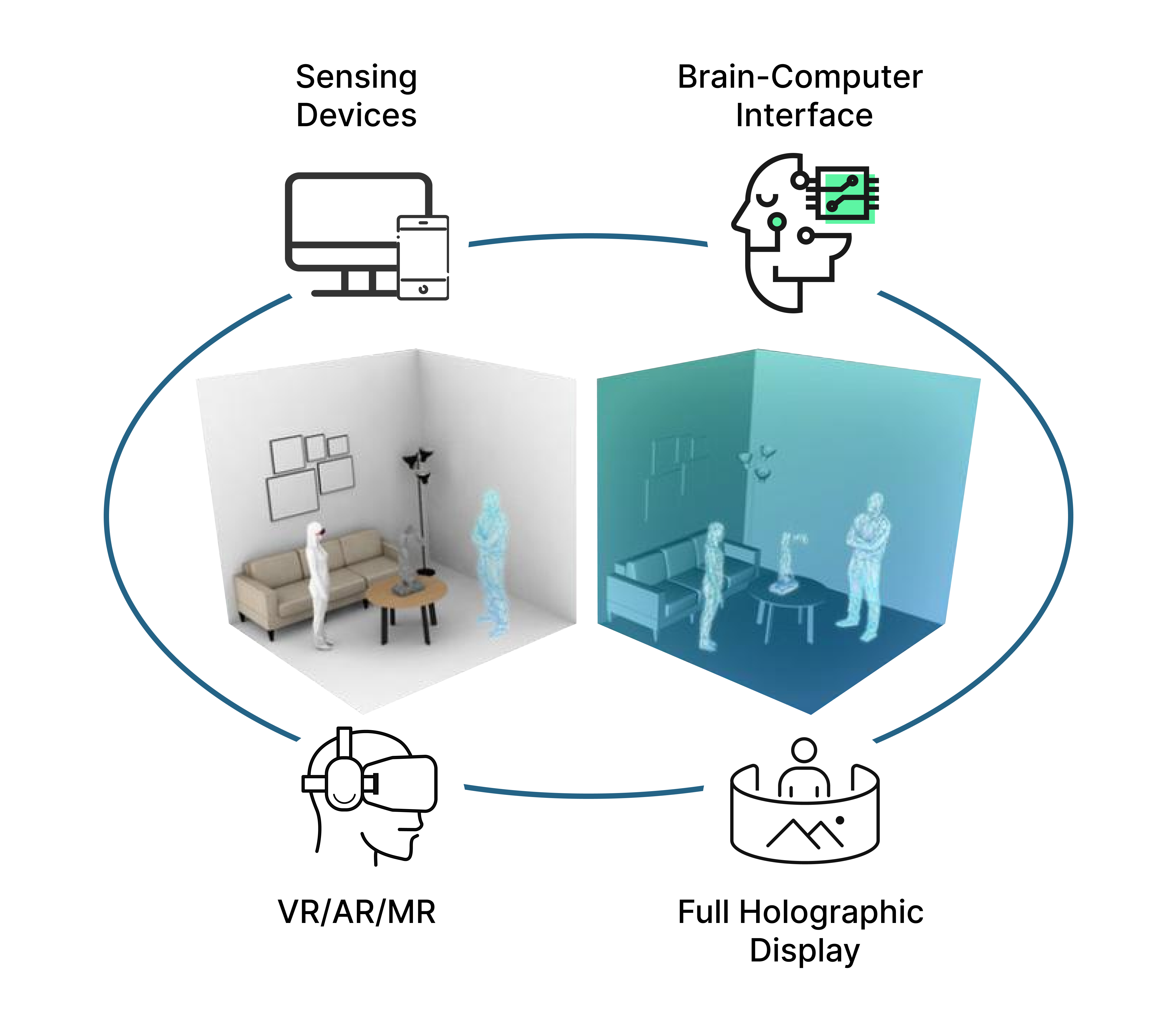}
    \caption{Compatibility Among Multiple Devices}
    \label{fig: multidevices}
\end{figure}

\begin{figure*}
    \centering
    \includegraphics[width=\textwidth]{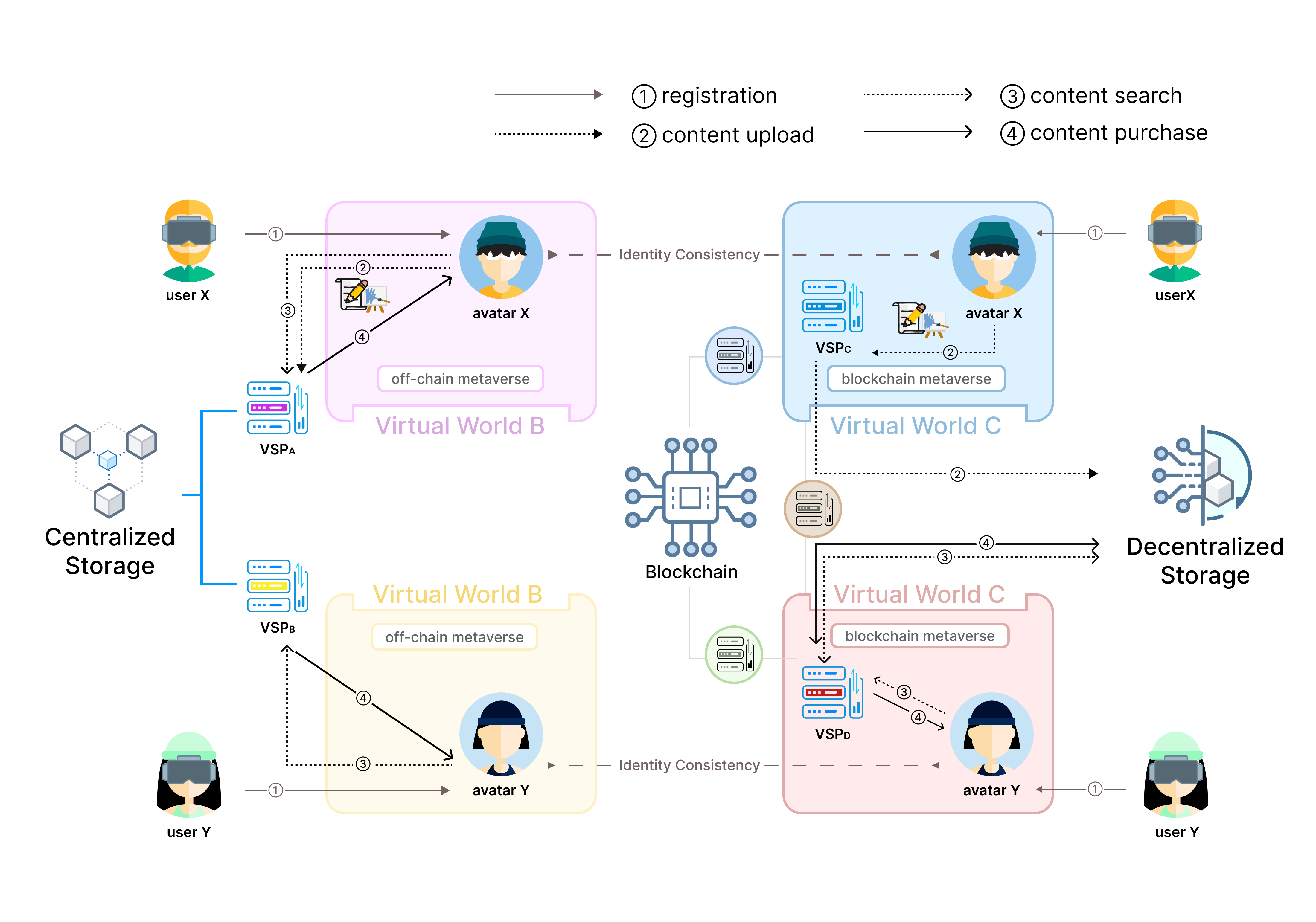}
    \caption{Seamless Navigation and Interoperability Among Platforms}
    \label{fig: multiplatforms}
\end{figure*}

We have systematically organized Metaverse Interoperability research into a structured framework comprising three following distinct layers, each encapsulating different dimensions, as depicted in Figure \ref{fig: full map}:


\begin{enumerate}
  \item \textit{Layer 1-Compatibility Among Multiple Devices}: This layer emphasizes interoperability in user experiences through VR, AR, and MR technologies, enabling seamless interaction across various smart device platforms, including personal computers and mobile phones. It also extends to emerging devices like brain-computer interfaces and holographic reality, as illustrated in Figure \ref{fig: multidevices}. 
  
  \item \textit{Layer 2-Seamless Navigation and Interoperability Among Platforms}: This layer focuses on the seamless navigation and transition of users' avatars and digital assets across different Metaverse environments. It leverages open data and technologies such as blockchain to facilitate the uninterrupted transfer of identities, assets, and user attributes, paralleling navigational experiences in the physical world, as illustrated in Figure \ref{fig: multiplatforms}.
  
  \item \textit{Layer 3-Integrated Interaction Between Physical and Virtual Worlds}: This layer addresses the convergence of physical and digital spaces. It includes capturing and reflecting an individual's expressions and movements onto an avatar with advanced technologies, and synchronizing physical and digital objects using IoT and Digital Twins, aiming for a unified interactive experience that seamlessly merges both worlds, as illustrated in Figure \ref{fig: physicalvirtual}.
  
\end{enumerate}

\begin{figure*}
    \centering
    \includegraphics[width=\textwidth]{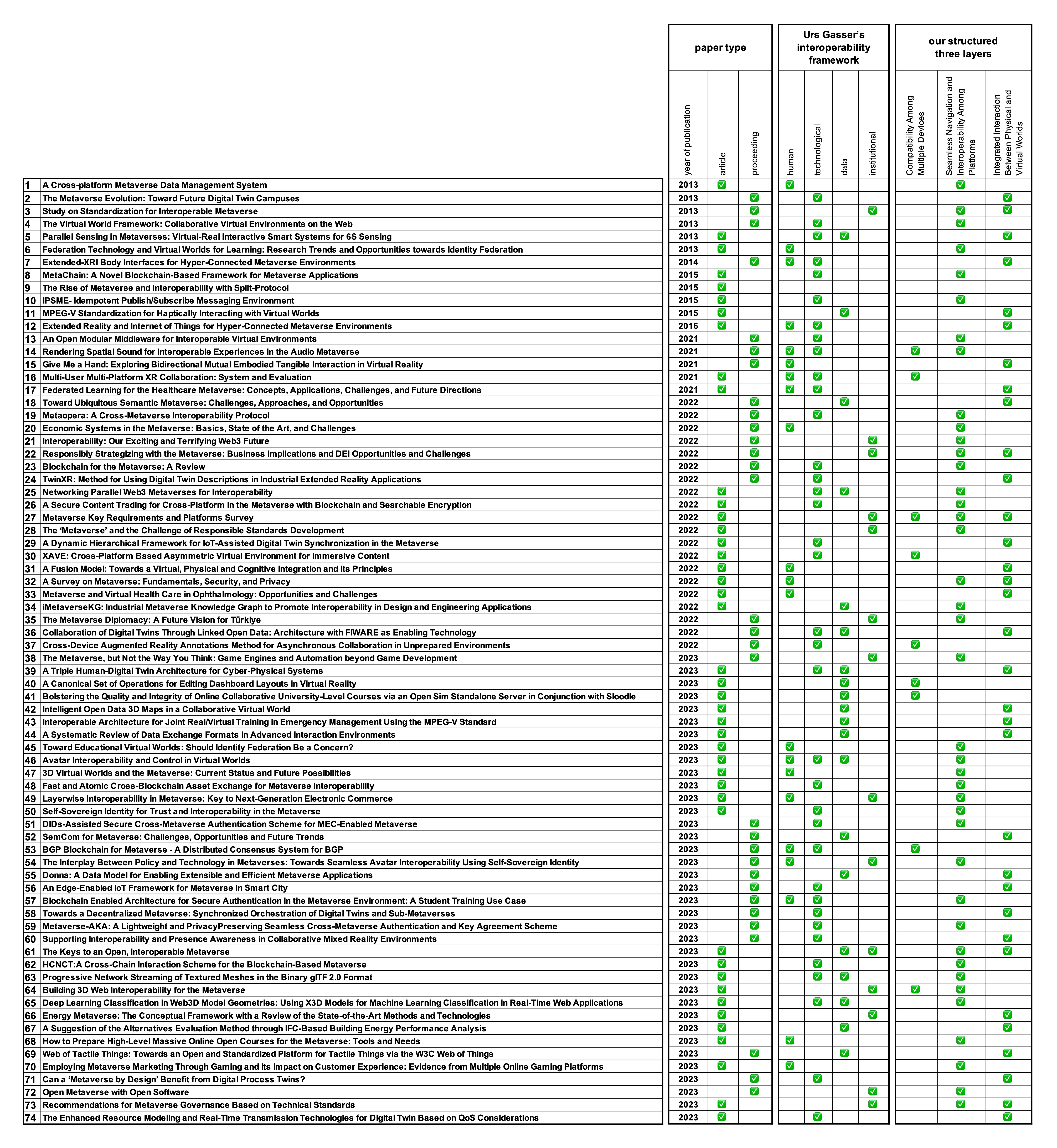}
    \caption{Literature Distribution Across Four Dimensions and Three Layers}
    \label{fig:overviewtable}
\end{figure*}

Combining Urs Gasser’s interoperability framework~\cite{gasser2015intero} and our three-layer Metaverse Interoperability with the core interoperability content extracted through content analysis, we have meticulously classified the core literature, given in Figure~\ref{fig:overviewtable}.

\subsection{The Human Dimension}

The human dimension refers to the capacity of individuals to comprehend and utilize the data exchanged, as well as their readiness to cooperate. Although it is more abstract than the technical and data layers, this layer can be conceptualized in terms of users' experiential needs and their acceptance of the experience during use. These factors have a direct influence on user satisfaction and the efficacy of system utilization~\cite{gasser2015intero}. 

From this dimension, seamless interaction is imperative across three layers: devices, platforms, and virtual-physical intersections.  The Metaverse, as an evolving digital ecosystem, should transcend three-dimensional technologies like VR, AR, and MR and bridge existing two-dimensional media with cutting-edge devices such as holographic displays and brain-computer interfaces, ensuring technological compatibility and operational continuity. Interoperability should enable smooth transitions for users among diverse virtual spaces, regardless of differences in software platforms, manufacturers, or countries, both functionally and experientially. As the Metaverse expands, it ought to enable users to navigate effortlessly between various services and platforms while safeguarding their data, privacy, and economic interests without disrupting the user experience. The blurring lines between physical and virtual worlds underscore the need for seamless integration, allowing users to transition and interact between these spheres without friction.

\subsubsection{Layer 1-Prioritize Cross-Device Experience}
Research on device integration and user preferences has revealed a significant demand for seamless cross-device experiences. Tumler et al.\cite{tumler2022multi} conducted preliminary studies into various combinations of xR devices to explore user requirements and preferences in cross-device operability. The results indicate that the integration of VR with personal computers and Hololens 2 is more preferred compared to other device combinations, highlighting a real demand for seamless cross-device experiences. Another study\cite{mike2023bgpblock} examines the paramount importance of interoperable architecture, asserting that user experience is a fundamental component of interoperability.

\subsubsection{Layer 2-Smooth Cross-Platform Navigation}
In an open, well-designed metaverse, individuals and entities and create diverse platforms with personalized avatars and environments for immersive experiences~\cite{Huang2023EconomicSys,shakila2023layerwiseI,siem2022selfsovereign}. The architecture enables seamless transitions among platforms, securely connecting user data for cross-platform interoperability and enhancing cultural and economic exchanges~\cite{Huang2023EconomicSys,Yfang2023NetWeb3,Jihyeon2023ASCTrading}. Users can expect their virtual possessions to be accessible or transferable without substantial change, mirroring the constancy we experience with physical belongings~\cite{dionisio20133d}. In an ideal Metaverse, such inherent continuity is fundamental. Hashem et al. found a significant correlation(r=0.73) between interoperability in this layer—the ability to manage virtual environments and assets—and user satisfaction through a quantitative analysis of survey data from 450 massively multiplayer online gamers~\cite{Hashem2023MartetingCE}. Zaman et al. identified key interoperability elements, such as consistent avatars, linked virtual identities, and accessible services, emphasizing global data sharing and digital asset interoperability~\cite{shakila2023layerwiseI}. Chi et al. discussed the interoperability of identity and assets, allowing recognition and liquidity across diverse environments, and the interoperability of behaviors and relationships, ensuring continuity and personalized interaction~\cite{Yfang2023NetWeb3}.

\subsubsection{Layer 3-Seamless Physical-Virtual Integration}
As technology advances, the importance of fluidity across the physical and virtual realms becomes paramount. While virtuality pres possibilities, our reliance on the physical world persists~\cite{Jie2022ERIoT}. To maintain its integrity and user retention, the Metaverse must cultivate a sense of presence and sustained engagement, enabling users to transition effortlessly between the two worlds~\cite{lal2021GiveH}. Achieving this requires the Metaverse to not only mirror every physical object in the virtual realm but also ensure their states are updated in real-time to prevent a fragmented user experience~\cite{ali2023federatedAI}. A lack of synchrony in information transfer between the physical and virtual worlds can hinder immersion, resulting in user fatigue and reduced engagement~\cite{Jie2022ERIoT}. Jie et al. explore the integration of digital content with the physical world using Internet of Things (IoT) technology to create more intelligent environments and enhance information flow~\cite{Jie2022ERIoT,Jie2022EXRI}. Similarly, Antonijevic et al. propose the use of IoT and 3D modeling to develop digital twins, effectively merging the real and digital realms~\cite{Petar2022MetaEvol}. Real-time synchronization is emphasized in healthcare~\cite{ali2023federatedAI,Tan2022Healthcare} and education~\cite{Petar2022MetaEvol}. Bozgeyikli incorporates physical objects into VR for realistic interactions and immersion~\cite{lal2021GiveH}, while Zhang et al. propose a Fusion Universe model combining virtual, physical, and cognitive spaces adhering to physical laws~\cite{Hao2023FusionModel}. Seamless integration allows for a unified experience and user engagement in the Metaverse.

\subsubsection{Layer 1\&2\&3- Emerging Metaverse Identity Challenges}

Our literature review identifies identity as the cornerstone of Metaverse interoperability. Digital identities, representing individuals, institutions, or objects with their attributes and preferences~\cite{Anwar2006Privacy}, are essential for distinguishing users and assets across Metaverse environments~\cite{Tao2023Metaopera,venugopal2023realm}. Previous studies highlight federated identity solutions as crucial for effective user and identity management~\cite{Goncalo2013FederationIDVW,Cruz2015EVWIdenFeder}. Identity federation facilitates secure information sharing between organizations, relying on robust identity management, security, and trust. Integrating an independent federated identity system into the infrastructure is recommended~\cite{Cruz2015EVWIdenFeder,Goncalo2013FederationIDVW,Cruz2015EVWIdenFeder}. Patwe and Mane discuss the challenges of uniform authentication across educational sub-Metaverses and advocate for a unified identity system to support collaborative activities, cross-institutional learning, and resource access~\cite{Sonali2023BCAuthenti}.

Challenges also arise in developing, reusing, and ensuring the interoperability of 3D resources, alongside managing copyright and licensing, mitigating reputational risks, and navigating the ethical and legal complexities associated with avatar misconduct such as bullying and sexual harassment~\cite{Cruz2015EVWIdenFeder}. Recent studies further address these concerns. Iacono and Vercelli highlight the negative impact of sexual harassment on user well-being and security~\cite{Iacono2023EduMeta}. Lee et al. examined the impact of such misconduct on user self-presence, noting that while regulation solutions help identify and prevent risks, challenges persist~\cite{lee2023legal}. Hence the need for a unique user identity to bolster safety and ensure consistency~\cite{Iacono2023EduMeta}. Venugopal et al. advocate for robust digital identity management to enable avatar interoperability and service delivery, outlining centralized, federated, and self-sovereign identity models~\cite{venugopal2023realm}. Laborade et al. propose using Self-Sovereign Identity (SSI) with offline governance protocols to ensure seamless interoperability, data privacy, and attribute portability~\cite{romain2023interplay}. Effective identity management is paramount for Metaverse interoperability, fostering trust and paving the way for an integrated ecosystem. Although this field has seen increasing interest~\cite{Goncalo2013FederationIDVW,Tao2023Metaopera,Yfang2023NetWeb3,wang2023survey,shakila2023layerwiseI,ying2023DIDs,romain2023interplay,Sonali2023BCAuthenti}, it is still in its nascent stages and requires further research to reach its full potential.

\subsection{The Technological Dimension}
The technical dimension pertains to the underlying infrastructure technology that facilitates system interconnection and data sharing, typically through predefined interface technologies~\cite{gasser2015intero}.

\subsubsection{Layer 1-Exploring Secure and Interoperable Architecture}
To realize the vision of Metaverse, reducing techninological barriers in the access infrastructure is essential for widespread interoperability. This foundational interoperability enables seamless interactions across systems and devices. Such infrastructure should allow universal access through various electronic devices and support collaborative experiences. Despite the lack of a unified software framework for all devices~\cite{tumler2022multi}, research is advancing. Tümle et al. have developed a client-server model using standard xR SDKs to enable multi-user experiences across diverse platforms~\cite{tumler2022multi}. Cho et al. introduced the XAVE architecture, integrating non-immersive and immersive experiences within heterogeneous virtual environments~\cite{yunsik2023xave}. This architecture supports devices from PCs and mobile devices to VR/AR headsets and motion capture systems, facilitating cross-platform interactions through keyboards, mice, touchscreens, game controllers, and image recognition sensors. Effective synchronization and network data management are crucial for its success. Beyond device interoperability, universal compatibility within device categories is equally important. For instance, a new AR collaboration method allows phones with varying performance levels to asynchronously generate and display AR annotations, ensuring compatibility despite performance differences~\cite{garcia2021cross}. Consistency in auditory and haptic experiences is also critical; Jot et al. developed a 6-DoF audio engine for the Metaverse that synchronizes audio and visuals, merges precomputed and real-time acoustic simulations, and supports cross-platform development with an open scene description model~\cite{jean2021rspatial}. Additionally, integrating the Border Gateway Protocol (BGP) with Distributed Consensus Systems (DCS) is being explored to create a secure and interoperable Metaverse architecture~\cite{mike2023bgpblock}.


\subsubsection{Layer 2-From Middleware to Blockchain}

In pioneering interoperability research among platforms, Byelozyorov et al. developed a modular, open-source middleware that uses common interfaces to connect different virtual worlds. This design supports dynamic linking of client and server interfaces, establishing principles for Interface Definition Language (IDL), APIs, communication protocols, and transportation mechanisms. Embedded compilers and interpreters enhance data processing and portability, successfully connecting previously incompatible platforms~\cite{Sergiy2013OpenMiddleware}. Burn et al. introduced the Virtual World Framework (VWF), a web-based 3D multi-user application framework using WebGL and WebSockets to promote interoperability among virtual worlds~\cite{Eric2014VWFramwork}. Preda and Jovanova engineered a system allowing avatars to move seamlessly between virtual worlds, covering geometry, appearance, animations, and attributes. This system balances avatar uniqueness with compatibility, addressing complexities like mesh and texture resolutions, and ensuring acceptance by different virtual world creators~\cite{Preda2013AvatarIntero}.OpenSimulator and its extension Hypergrid have been widely adopted in the industrial sector for creating and managing custom 3D environments, promoting user and data interoperability across various OpenSim virtual worlds~\cite{Cruz2015EVWIdenFeder}. 

These early efforts laid the groundwork for the ongoing evolution of metaverse interoperability. Recent advancements include the IPSME architecture by Nevelsteen and Wehlou, which integrates disparate systems using a publish-subscribe mechanism and dynamic translators. This approach supports system evolution and simplifies integration without requiring uniform protocols, and has been successfully applied in scenarios like a Minecraft metaverse instance~\cite{Kim2021IPSME}. Chen et al. proposed the Cross-Platform Metaverse Data Management System (CMDMS), enabling users to access profiles and spaces across platforms and facilitating the transfer of 3D digital assets between Unity and Unreal engines~\cite{Bohan2022Acrossplat}.


Recent literature categorizes the Metaverse into centralized and decentralized frameworks. Centralized metaverses operate on central servers managed by single entities. In contrast, decentralized metaverses use blockchain technology to enable distributed management, user asset ownership, autonomous trading, and transparent self-governance~\cite{Tao2023Metaopera,Huang2023EconomicSys,Yfang2023NetWeb3,Akbobek2023MetaPlatform}. Research on Metaverse interoperability focuses on interactions within centralized metaverses, within decentralized metaverses, and between the two. Technologies for centralized metaverses interoperability mirror traditional virtual worlds, though detailed discussions are limited~\cite{Tao2023Metaopera,Akbobek2023MetaPlatform}. Decentralized metaverses leverage cross-chain technology for asset and data interchange across blockchain platforms~\cite{Tao2023Metaopera,Huang2023EconomicSys,Yfang2023NetWeb3}. 
Interoperability between centralized and decentralized metaverses requires complex on-chain and off-chain technological collaboration~\cite{Tao2023Metaopera}. Chi et al. identify four types of decentralized metaverse interoperability: within the same project across different blockchains via centralized servers or cross-chain bridges; among different projects on one blockchain through token swaps on decentralized exchanges; across multiple blockchains using centralized exchanges or cross-chain technologies; and within one project on a single blockchain via native on-chain protocols~\cite{Yfang2023NetWeb3}. Li et al. further classify interoperability into cross-chain interactions within decentralized metaverses and integrations between decentralized and centralized metaverses, both on-chain and off-chain. They propose the MetaOpera protocol, which significantly improves existing solutions by reducing transaction proof sizes eightfold and enhancing latency threefold~\cite{Tao2023Metaopera}.


Our literature review highlights blockchain technology as the cornerstone of metaverse interoperability. It is crucial for identity management~\cite{Gadekallu2022BlockchainFT,Mohammad2023Metaverse,Wang2023ARO,Grner2018OnTR}, security~\cite{Jongseok2022DesignBC,Gadekallu2022BlockchainFT,Singh2023Enhance,Jamil2023BlockchainBasedDF}, asset protection~\cite{Truong2023BlockchainMM,Ersoy2022BlockchainbasedAS}, and data preservation~\cite{Jongseok2022DesignBC,GA2022STUDYOB,Shilpi2022ContributionBC}, while also facilitating integration with real-world economies~\cite{Jayandren2022BeyondXRBC}. Blockchain enables seamless asset transactions across virtual domains, securing identities and assets, and supports a decentralized, open-source metaverse conducive to application development and digital commerce~\cite{Huang2023EconomicSys,huynh2023blockchain}. Extensive research explores these facets. \textit{For identity management}, Chirmai et al. propose integrating Self-Sovereign Identity (SSI) with blockchain to enhance security and interoperability~\cite{siem2022selfsovereign}. Yao et al. introduce an architecture combining multi-access edge computing with blockchain, using Decentralized Identifiers (DIDs) to improve authentication security and reduce resource usage~\cite{ying2023DIDs}. Patwe and Mane develop a blockchain-based method to protect educational metaverse identities from impersonation and attacks~\cite{Sonali2023BCAuthenti}, while Yao et al. devise a Metaverse-AKA authentication system for privacy and cross-metaverse authentication~\cite{Ying2022AuthKeyAgree}. \textit{For asset security}, Jihyeon Oh et al. create a blockchain-based content trading system enhancing transparency and security across metaverse platforms~\cite{Jihyeon2023ASCTrading}. Jiang et al. develop an efficient cross-blockchain asset exchange protocol for rapid trading~\cite{jiang2023fast}, and Ren et al. propose a cross-chain transaction strategy using an improved Hashed Timelock Contract (HTLC) mechanism to mitigate risks and reduce centralization~\cite{Yongjun2023HcnctBC}. \textit{In data management}, blockchain facilitates data exchange with cross-chain protocols~\cite{huynh2023blockchain}. \textit{For resource management and incentivization}, Nguyen et al. construct the MetaChain framework using blockchain and smart contracts to enhance provider-user interactions and employ game theory for incentivization~\cite{Cong2022ANovel}. These studies underscore blockchain's role in promoting interoperability and security in the metaverse. However, cross-metaverse interoperability research is still nascent and requires further exploration~\cite{Tao2023Metaopera,huynh2023blockchain,Huang2023EconomicSys}.


\subsubsection{Layer 3-Technologies Bridging Physical-Virtual Worlds}

\begin{figure}
    \centering
    \includegraphics[width=0.65\linewidth]{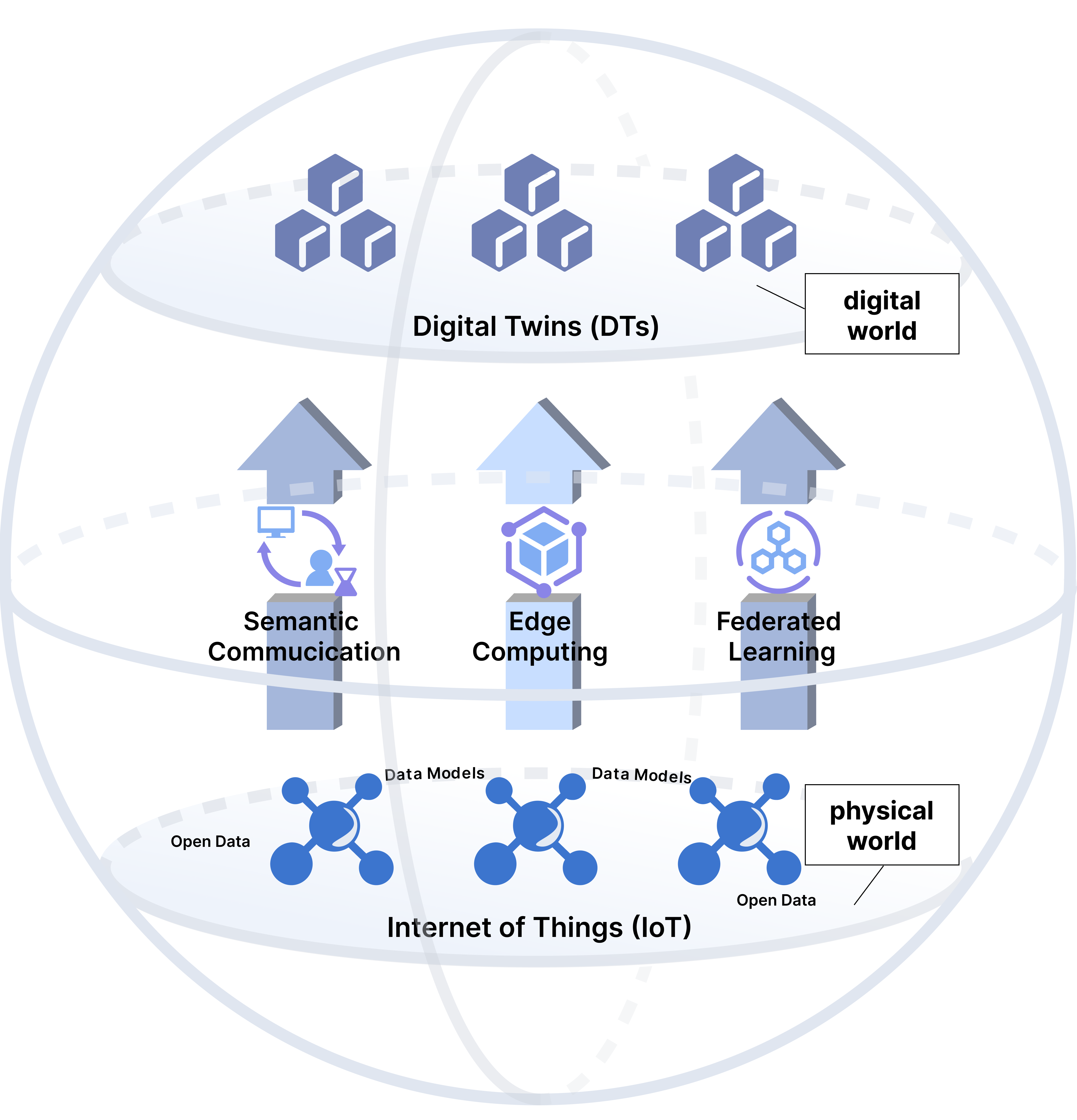}
    \caption{Integrated Interaction Between Physical and Virtual Worlds}
    \label{fig: physicalvirtual}
\end{figure}

The technological dimension here refers to the essential technologies that enable seamless interaction and data exchange between the real and virtual worlds. These technologies ensure real-time reflection of data and states between both realms. First, extensive real-world information is collected via sensors, including environmental factors and behaviors~\cite{Petar2022MetaEvol,Jie2022ERIoT}. This data is then processed by digital twins to enhance the virtual environment~\cite{Petar2022MetaEvol}. Additionally, the technological layer secures data through protective mechanisms, maintaining privacy and efficiency~\cite{ali2023federatedAI,Lim2023MetaEdgeI}. Our review highlights three key areas: (1) Sensors and IoT; (2) Digital Twins; and (3) Edge Computing, Federated Learning, and Semantic Communication, as illustrated in Figure \ref{fig: physicalvirtual}.


\paragraph{Sensors and Internet of Things(IoTs)}
Sensors and IoTs are crucial in the phygital Metaverse, transmitting information via VR, AR, and MR headsets embedded with sensors that track head movements. These sensors synchronize the virtual environment with the user's physical position, enhancing immersion. They capture signals from users and their environments, improving interactivity and naturalness. Sensors can be centralized in headsets or distributed throughout the environment, facilitating wider interaction and data acquisition. This integration aligns VR, AR, and MR technologies with IoT. Edge devices in IoT sense, communicate, and respond, while in MR, they make virtual interactions more lifelike and instinctive. Strategic sensor placement ensures seamless data transmission between physical and virtual realms~\cite{Jie2022EXRI}. Yue notes that during the Metaverse's early stages, physical asset owners may not fully integrate or support interoperability standards, making IoT devices essential for real-time status updates~\cite{Yue2023ADynamicF}. Bashir et al. stress the need for comprehensive mapping of real-world objects and instant updates to their virtual counterparts for a coherent user experience~\cite{ali2023federatedAI}. Shen et al. introduced "Parallel Sensing," combining physical sensors with digital twins to enhance perception capabilities and compensate for intermittent physical sensor operation~\cite{yu2022ParallelSensing}. Lee et al. show that adopting oneM2M standards with edge computing accelerates data transfer, enhances processing precision, and improves real-time linkage between physical objects and the Metaverse, boosting interoperability and user experience~\cite{Jiho2023AnIoT}.


\paragraph{Digital Twins(DTs)}
Digital Twins (DTs) create precise digital replicas of physical entities, dynamically mirroring their structures, states, and behaviors in real-time~\cite{Yue2023ADynamicF}. Hashash et al.'s distributed Metaverse framework exemplifies this by synchronizing physical and digital twins via Mobile Edge Computing (MEC)\cite{Omar2023SynDTSubM}. Han et al. highlight DTs' role in enhancing Metaverse interoperability among Virtual Service Providers (VSPs), proposing a dynamic framework for improved synchronization\cite{Yue2023ADynamicF}. Shangguan et al. introduced a triadic architecture for interactions among humans, objects, and DTs, demonstrated in lunar rover power management~\cite{shangguan2022triple}.
Unified standards for data modeling, representation, and communication in DTs are essential for widespread Metaverse interoperability. Li et al. propose a framework for improved DT interoperability through advanced modeling and real-time transmission~\cite{Li2022enhancedDTs}. Conde et al. integrate FIWARE and Linked Open Data for effective DT communication in urban settings~\cite{Javier2022DTsOpenData}. The BEAMING project by Oyekoya et al. supports efficient multi-device communication with minimal latency~\cite{Oyewole2013InterMR}.
Integrating DTs with other technologies is crucial for expanding Metaverse applications. Stary's research on digital process twins enhances architectural design interoperability~\cite{Stary2023DTs}. Tu et al.'s TwinXR project merges DTs with Extended Reality (XR) for bidirectional data flow and system interoperability, validated in smart manufacturing~\cite{tu2023twinxr}. Despite advancements, challenges in real-time DT synchronization persist, requiring efficient computing and connectivity~\cite{Omar2023SynDTSubM}.


\paragraph{Edge Computing, Semantic Communication, and Federated Learning}

Edge Computing (EC) enhances efficiency and reduces latency by processing data close to its source, crucial for Metaverse interoperability. Hashash et al. and Lee et al. show EC significantly decreases synchronization latency between the real world and digital twins, essential for real-time user experiences~\cite{Omar2023SynDTSubM,Jiho2023AnIoT}. Hashash et al.'s framework, using edge-based reinforcement learning, cuts synchronization times by 25.75\% and improves inter-system connectivity~\cite{Omar2023SynDTSubM}. Lee et al. find that EC, combined with AI, accelerates data transfers and enhances interactions between smart cities and virtual environments, supporting smart city development and IoT devices~\cite{Jiho2023AnIoT}.

Semantic Communication (SemCom) ensures interoperability by extracting and transmitting only meaningful data, reducing volume, bandwidth needs, and latency~\cite{Georgios2023Donna,Saeed2023SemCom}. SemCom efficiently manages extensive human-centric Metaverse data~\cite{Saeed2023SemCom}. Bouloukakis and Kattepur highlight its role in scalability and interoperability through semantic mapping, standard data models, and interaction capture. Accurate semantic descriptions are crucial for correct data capture and transmission~\cite{Georgios2023Donna}. Li et al. propose a framework with AI, Spatio-Temporal Data Representation (STDR), Semantic IoT (SIoT), and Semantic-enhanced Digital Twins (SDT) to optimize data transmission and conserve bandwidth~\cite{Kai2023UbiquitousSemantic}. Future research should focus on cross-domain semantic mapping to enhance interoperability~\cite{Saeed2023SemCom}.

Federated Learning (FL) is key for Metaverse security and privacy, allowing machine learning on distributed datasets without exposing raw data~\cite{Saeed2023SemCom,ali2023federatedAI}. Li et al. advocate FL for protecting semantic data privacy~\cite{Saeed2023SemCom}. Bashir et al. explore FL's potential in the medical Metaverse, where it maintains data confidentiality and aids disease modeling and trend analysis despite data heterogeneity. Vertical FL enhances interoperability for datasets with shared identifiers but different features, supporting intelligent healthcare solutions~\cite{ali2023federatedAI}.

\subsection{The Data Dimension}
The Data Layer pertains to the capability of interconnected systems to comprehend mutually, encompassing aspects such as data formats, structures, semantics, and protocols. This is essential for ensuring information is transmitted and interpreted accurately between disparate systems. It is intrinsically associated with the Technology Layer and is frequently maintained for conjunction~\cite{gasser2015intero}. 

\subsubsection{Layer 1-Data Format Structures Among Multiple Devices}
Our literature review, which emphasizes the broader constructs of the Virtual Worlds and the Metaverse, has yielded limited discourse on specific data format structures, attributable to a non-device-specific focus. Within the selected body of work, two pivotal studies address the challenges of interoperability between 2D and 3D data content. These investigations delve into integrating content from established 2D digital ecosystems into the nascent 3D environments of the Metaverse. Such integration not only augments the content base, mitigating the scarcity of content in the initial phases of the Metaverse development~\cite{Setti2021twoD3D}, but also ensures that the development of the 3D Metaverse is in alignment with the established 2D digital milieu, considering user experience requirements~\cite{Pellas2016VWvLMS}. Current research is examining the methods by which 2D content may be invoked and structured within 3D spaces~\cite{Setti2021twoD3D}, as well as how 3D virtual environments can interconnect with conventional 2D learning management systems~\cite{Pellas2016VWvLMS}. These endeavors are instrumental in forming the foundational elements of the Metaverse technology's advancement toward seamless interoperability.


\subsubsection{Layer 2-Data Formats Standardization Among Platforms}

Seamless navigation among the various platforms hinges on data interoperability, a critical cornerstone. Such interoperability typically requires adopting a universal data protocol to ensure that information can be seamlessly transmitted at the base level and interpreted accurately. 3D Data format standardization is crucial for Metaverse interoperability in this dimension, enabling seamless data exchange across diverse systems~\cite{Santo2015DataExInter}. Standardized formats facilitate accurate data transfer, reduce technical barriers in cross-system interactions, and empower developers to create cross-platform tools efficiently~\cite{Chrysoula2022X3D}. This promotes innovation, collaboration, and enhances AI-driven data analysis, leading to smarter, personalized Metaverse experiences.

VRML (Virtual Reality Modeling Language)~\cite{vrml}, established in 1994 (ISO/IEC 14772-1:1997), was an early 3D graphics format for detailed, interactive environments. Despite its pioneering role, VRML faced adoption issues due to performance problems, limited browser support, and a complex interface~\cite{Preda2013AvatarIntero,dionisio20133d,Lim2023MetaEdgeI}.

X3D (eXtensible 3D)~\cite{x3d}, the successor to VRML, became the international standard for 3D graphics content (ISO/IEC 19775-19777) starting in 2005. X3D offers enhanced graphics, XML-compatible syntax, and supports various encoding formats, promoting interoperability and accessibility. The latest version, X3D4, integrates HTML5, advanced rendering, and WebAudio API, supporting cross-domain data integration and realistic rendering~\cite{Anita2022X3D,web3d_x3d}. X3D's extensibility and compatibility with semantic web technologies facilitate innovative, real-time 3D applications~\cite{Chrysoula2022X3D}.

glTF (GL Transmission Format) 2.0~\cite{khronos_gltf}, recognized by the Metaverse Standards Forum (MSF) as a fundamental standard for 3D asset interoperability, was developed by the Khronos Group in 2017. Known as the "JPEG of 3D," glTF 2.0 optimizes 3D model transmission for efficient content delivery. Its JSON-based structure and binary storage enhance performance and reduce file sizes, supporting realistic visual effects and complex animations~\cite{Wouter2023glTF,khronos_gltf}. Major 3D platforms like Unity3D and Blender support glTF 2.0.

USD (Universal Scene Description)~\cite{usd}, developed by Pixar and open-sourced in 2016, excels in managing complex scenes in collaborative environments. It supports non-destructive editing, external references, and relational scene elements, facilitating efficient data interchange across software~\cite{usd}. USD's growing industry adoption and recognition by the MSF highlight its importance for 3D asset interoperability~\cite{aousd_core_initiatives,msf2023}.

COLLADA (Collaborative Design Activity)~\cite{collada} is an XML-based format designed for neutral 3D asset exchange among graphics applications. Unlike VRML, X3D, or glTF, COLLADA focuses on interoperability rather than comprehensive 3D visualization, ensuring smooth file transfers across software platforms~\cite{dionisio20133d}. 

Together, these formats represent the current leading 3D data format standardization in the metaverse, each playing a crucial role in the exchange of data and the development of applications. Table~\ref{tab:3d_format_comparison} is constructed to compare their design goals and features succinctly.
\begin{table*}[t]\scriptsize
\centering
\caption{Comparison of 3D Data Formats in the Metaverse}
\label{tab:3d_format_comparison}
\begin{tabular}{m{1.6cm} m{1.6cm} m{1.6cm} m{2.1cm} m{2cm} m{2cm}}
\toprule
\multicolumn{1}{c}{\textbf{Format}} & \multicolumn{1}{c}{\textbf{VRML}} & \multicolumn{1}{c}{\textbf{X3D}} & \multicolumn{1}{c}{\textbf{glTF 2.0}} & \multicolumn{1}{c}{\textbf{USD}} & \multicolumn{1}{c}{\textbf{COLLADA}} \\ \midrule
\textbf{Design Goal}                         & 3D visualization                  & 3D visualization                 & Real-time rendering                   & Complex scenes, \mbox{animations}       & Interoperable asset \mbox{exchange}         \\ \hline
\textbf{Optimized For}                       & Visualization                     & Visualization                    & Web display                           & Large-scale data \mbox{management}      & Flexible content \mbox{exchange}            \\ \hline
\textbf{Interoperability}                    & Moderate                          & High                             & High                                  & High                             & High                                 \\ \hline
\textbf{Flexibility}                         & Moderate                          & High                             & High                                  & Very High                        & High                                 \\ \hline
\textbf{Performance}                         & Low                               & Moderate                         & High                                  & Very High                        & Moderate                             \\ \hline
\textbf{Industry Adoption}                   & Low                               & Moderate                         & High                                  & Increasing                       & Moderate                             \\ \hline
\textbf{Software-Agnostic}                   & No                                & No                               & Yes                                   & Yes                              & Yes           \\ \bottomrule 
\end{tabular}
\end{table*}

\subsubsection{Layer 3-Data Formats Standardization Between Physical and Virtual Worlds}
In this layer, key data format standards include ISO/IEC 23005 (MPEG-V)~\cite{mpegv} and IEEE 2888~\cite{ieee2888}. MPEG-V bridges virtual and physical environments, managing sensory feedback like vision, sound, and touch, and includes architecture, object encoding, and exchange protocols~\cite{wang2023survey, Jaeha2012MPEGV, Ardila2015TrainMPEGV, Santo2015DataExInter}. It enables users to interact with virtual spaces using sensory devices and brings virtual effects into the real world~\cite{mpegv, Santo2015DataExInter}. IEEE 2888, introduced in 2019, complements MPEG-V by standardizing data formats and APIs for synchronizing virtual and physical worlds, enhancing sensor data acquisition and actuator control~\cite{ieee2888}. Kim et al. highlight MPEG-V's role in haptic technology, showing its application in editing and presenting haptic content~\cite{Jaeha2012MPEGV}. Ardila et al. extend MPEG-V to a real-time training framework, improving interoperability through an enhanced data model~\cite{Ardila2015TrainMPEGV}. 

Beyond MPEG-V, innovative approaches like the World of Tactile Things (WoTT) are emerging. WoTT, proposed by Pham et al., aims to create a tactile Internet and standardize tactile data exchange using W3C's Web of Things standards. It enables effective information exchange between tactile sensing devices across different domains, facilitating the creation of digital twins~\cite{pham2022tactile}. Additionally, oneM2M, a global IoT interoperability standard, supports end-to-end connectivity for IoT services and applications. This convergence of IoT and the Metaverse positions oneM2M as essential for developing a unified virtual-physical environment~\cite{Jiho2023AnIoT}. IEEE 2888, introduced in 2019, complements MPEG-V by standardizing interfaces for synchronizing the virtual and physical worlds. It specifies data formats and APIs for accurate sensor data acquisition and actuator control, advancing the integration of virtual-physical systems~\cite{ieee2888}. Kim et al. provide an overview of the MPEG-V standard, highlighting the parts closely related to haptic technology and demonstrating how MPEG-V can be used in the editing, creation, and presentation of haptic content~\cite{Jaeha2012MPEGV}. Ardila et al. extend MPEG-V to a real-time training framework, enhancing interoperability through an improved MPEG-V-based data model, reinforcing real-virtual connections~\cite{Ardila2015TrainMPEGV}.

\subsubsection{Layer 1\&2\&3-Approaches beyond Data Standardization}

Data interoperability in the Metaverse is a complex challenge due to the wide range of data types and file formats. Different scenarios support various formats, creating a diverse data ecosystem. Developing a universal data standard is difficult due to distinct interactions and significant economic and technical hurdles in transferring data across platforms and scenarios. While a common protocol could theoretically solve these issues, its practical implementation is challenging. Therefore, we need to explore solutions beyond data format standardization.

\paragraph{KGs-Based Data Integration}
Chi et al. suggest that Knowledge Graphs (KGs) can address interoperability challenges by integrating data from various sources and formats~\cite{Yfang2023NetWeb3}. KGs map entities and their relationships graphically, enhancing semantic richness with attributes~\cite{Rin2022FAIRKG}. They use Semantic Web technologies like RDF and OWL to standardize data and enable queries. This approach improves data integration, search capabilities, and natural language processing. The IEEE Knowledge Graph Working Group established standards to oversee this process in 2019, including IEEE P2807, which outlines the framework for KGs, and IEEE P2807.1, which specifies technical details and performance metrics~\cite{ieee2807, Utkar2022iMetaKG}. Jaimini et al. applied KG technology to support interoperability in the industrial Metaverse, developing a prototype for design engineering applications~\cite{Utkar2022iMetaKG}. While KGs are promising for enhancing data interoperability, this research area is still nascent and requires substantial further investigation and development~\cite{Yfang2023NetWeb3, Utkar2022iMetaKG,tu2023twinxr}.


\paragraph{Data Model Development}
Data models standardize data structuring to ensure system interoperability~\cite{Javier2022DTsOpenData}. Shangguan et al. describe digital twin data as multi-layered, incorporating datasets like human behaviors, physical entities, and virtual models, enabling real-time data capture and dynamic two-way mapping between the virtual and physical domains~\cite{shangguan2022triple}. Despite advancements in network infrastructure, comprehensive modeling of physical-virtual interactions is still underexplored~\cite{Georgios2023Donna}. Bouloukakis and Kattepur introduced the Donna data model, which uses a property graph schema to detail interactions among spaces, sensors, devices, and human-avatar pairings. In a virtual museum, the Donna model manages perception, attribute updates, and semantic interactions, demonstrating its versatility~\cite{Georgios2023Donna}. As the field of semantic representation and shared modeling in the Metaverse grows, these models are vital for ensuring interoperability, scalability, and adaptability~\cite{Georgios2023Donna}.


\paragraph{Open Data Utilization}
Open data, often funded and provided by government agencies, includes geographic, meteorological, and transportation datasets~\cite{Virtanen2015threeDVM}. While not always in the public domain, it is accessible under open licenses that allow free use and repurposing~\cite{lee2017opendata}. Open data is crucial for interoperability, aiding in the creation of 3D models, digital twins, and the Metaverse~\cite{Virtanen2015threeDVM,Javier2022DTsOpenData}. Virtanen et al. used open data from Finland's National Land Survey to develop a 3D virtual mapping environment with real-time updates and collaborative features~\cite{Virtanen2015threeDVM}. To address interoperability in digital twins, Conde et al. proposed a communication mechanism using the FIWARE Data Model and Linked Open Data (LOD) for two-way information exchange~\cite{Javier2022DTsOpenData}. This ecosystem involves data publishers, platform maintainers, and users, with standardized formats like NGSI-LD and metadata standards such as DCAT being crucial for seamless integration. By embracing standardization and automation, open data enhances interoperability and drives innovation in the Metaverse.



\subsection{The Institutional Dimension}
In Gasser's framework, the institutional layer is defined as the capacity for effective participation within a social system, underscoring the significance of institutional dimensions in digital ecosystems' interoperability, often equal to or greater than technological considerations~\cite{gasser2015intero}. Park et al. highlight that institutional interoperability involves integration across organizations, while regulatory interoperability requires common governance and inclusive stakeholder participation~\cite{park2023Web3Intero}. Zaman et al. add that Metaverse interoperability should address governance/business, experience, content, and infrastructure layers, focusing on standards, policies, and regulations to protect intellectual property and ensure privacy and security~\cite{shakila2023layerwiseI}. Our discussion on institutional interoperability will cover three main areas: 1) Standards Development Organizations (SDOs), 2) industrial contributions, and 3) policy frameworks.


\subsubsection{Standards Development Organizations}
SDOs play a crucial role in establishing standards for interoperability, security, quality, and reliability across industries and technologies. These standards ensure seamless integration and reliable operation of diverse systems and products. Although research on SDO dynamics is emerging~\cite{Anita2022X3D,Anita20233DWebInter,wook2023standardization,Perri2023open,romain2023interplay,Akbobek2023MetaPlatform}, comprehensive analyses are limited. Table \ref{tab:SDOs} summarizes the current status and progress of key SDOs involved in Metaverse interoperability. Many SDOs are still in the early stages of addressing the Metaverse's complex and evolving standardization needs. Given the Metaverse's novelty, these organizations need to invest more in research, dialogue, and consensus-building. Their current efforts focus on defining fundamental concepts, creating foundational frameworks, and setting preliminary technical guidelines to promote structured growth in the sector. As the Metaverse evolves, the influence of SDOs is expected to grow significantly.


\begin{table*}[t]\scriptsize
\centering

\caption{Overview and Characterization of Key Metaverse SDOs (M\_I\_Repre: Representative standards for Metaverse Interoperability)}

\label{tab:SDOs}
\begin{tabular}{m{1.8cm} m{1.2cm} m{3.4cm} m{2.2cm} m{0.9cm} m{0.5cm} m{1cm}}
\toprule
\textbf{SDO} & \textbf{M\_I\_Repre} & \textbf{Mission} & \textbf{Membership} & \textbf{Influence} & \textbf{Year} & \textbf{Status} \\ \midrule

\textbf{W3C} & WebVR, \mbox{WebXR} & Web standards development for long-term growth & Organization \mbox{members} & High & 1994 & Active  \\ \hline

\textbf{Web3D \mbox{Consortium}} & X3D & Standardize web-based 3D graphics for seamless use and growth across devices and platforms & Organization \mbox{members} & High & 1997 & Active \\ \hline

\textbf{ISO/IEC MPEG} & MPEG-V & Develop international standards for the coding, compression, and transmission of audio, video, and related data & National-bodies \mbox{of ISO and IEC} & High & 1988 & Active \\ \hline

\textbf{Khronos Group} & glTF2.0, OpenXR & Develop advanced, dynamic, open, and royalty-free interoperability standards & Industry consortium & High & 2000 & Active  \\ \hline

\textbf{ITU-T \mbox{CG-Metaverse}} & - & Dedicate efforts explicitly toward Metaverse standardization. & Mainly Countries and Sector Members & High & 2012 & Developing \\ \hline

\textbf{Open \mbox{Metaverse} \mbox{Interoperability} Group} & - & Facilitate open interoperability in the Metaverse & Open to individuals and organizations & Growing & 2021 & Formative stages \\ \hline

\textbf{IEEE \mbox{Metaverse} \mbox{Standards} \mbox{Committee}} & P2048 \mbox{standards} et al. & Develop standards for metaverse,VR,AR and advocate them on global basis & Professional association members & Growing & 2022 & Active \\ \hline

\textbf{World \mbox{Metaverse} \mbox{Council}} & - & Provide guidance for global Metaverse policies and standards & Various stakeholders & Growing & 2022 & Developing \\ \hline

\textbf{Alliance \mbox{for OpenUSD}} & USD & Advancing interoperability in 3D content creation through USD & Pixar, Adobe, Apple, Autodesk, and others & Growing & 2023 & Active \\ \hline

\textbf{Metaverse Standards Forum} & - & Promote open standards and interoperability in the Metaverse & Over 1,200 organizations & Significant potential & 2022 & Active \\ \hline
\toprule
\end{tabular}
\end{table*}

\paragraph{Web3D Consortium (Web3D)} Founded in 1997, the Web3D Consortium~\footnote{https://www.web3d.org/} is an international, non-profit organization dedicated to standardizing 3D graphics technologies. This member-driven consortium develops royalty-free standards for the interactive and real-time exchange of 3D graphics on the Web. Key achievements include ISO-IEC standards like X3D (Extensible 3D)\cite{x3d} and H-Anim (Humanoid Animation)\cite{hanim}. X3D is notable for its openness, extensibility, interoperability, and platform independence, supporting 3D graphics on desktops, tablets, and smartphones~\cite{Anita2022X3D, Anita20233DWebInter}. At its October 2023 meeting, the consortium focused on the 3D Web interoperability working group's responsibilities and Metaverse interoperability challenges~\cite{Anita20233DWebInter}.

\paragraph{Khronos Group} Founded in 2000, the Khronos Group~\footnote{https://www.khronos.org/} is a non-profit consortium dedicated to developing open, royalty-free standards for cross-platform graphics and computing. It is known for standards like OpenGL, Vulkan, OpenCL, and WebGL. The consortium has significantly impacted Metaverse interoperability with standards such as glTF~\cite{khronos_gltf} and OpenXR~\cite{openxr}. OpenXR, in particular, provides a unified API for VR, AR, and MR applications across various platforms, allowing developers to deploy applications on any OpenXR-compliant device. This standard simplifies development and fosters innovation in cross-platform Extended Reality (XR) experiences. OpenXR covers essential functions like device management, scene composition, spatial tracking, and user interaction, making it a key enabler of XR technology interoperability and accelerating industry growth~\cite{wook2023standardization,Akbobek2023MetaPlatform,Perri2023open}.


\paragraph{W3C Metaverse-related WGs and CGs} The World Wide Web Consortium (W3C)~\footnote{https://www.w3.org/}, founded by Sir Tim Berners-Lee in 1994, develops Internet standards. Its mission is to create protocols and guidelines that enhance the Web's capabilities and ensure its long-term sustainability. Within W3C, Working Groups (WGs) and Community Groups (CGs) play key roles. CGs often develop specifications that inform WGs or lead to new WGs. Hyun identifies three W3C groups focusing on Metaverse interoperability. The Virtual Reality website Community Group and the Metaverse Community Group, both established in 2015, have shown limited activity recently. In contrast, the Galaxy Metaverse Community Group, initiated in January 2022, is actively working on infrastructure, land governance, marketing, avatar communication protocols, and commerce within virtual environments and metaverses~\cite{wook2023standardization}. Additionally, the Verifiable Credentials Working Group is advancing the W3C Verifiable Credentials and Decentralized Identifiers (DID) standards, crucial for identity verification and decentralized registry systems~\cite{romain2023interplay}.


\paragraph{ITU-T CG-Metaverse} The International Telecommunication Union (ITU)\footnote{https://www.itu.int/en/Pages/default.aspx\#/zh} has led global technological standardization for over 150 years. Within the ITU's Telecommunication Standardization Sector (ITU-T), efforts to standardize the Metaverse are concentrated in Study Groups SG16 (Multimedia), SG17 (Security), and SG20 (IoT and smart cities). In December 2022, ITU-T SG16 established the Focus Group on the Metaverse (FG-Metaverse)\cite{ITU} under the Telecommunication Standardization Advisory Group (TSAG). The FG-Metaverse is structured into Working Groups (WGs) and Task Groups (TGs), responsible for developing standards and facilitating discussions. Currently, FG-Metaverse has nine WGs, nineteen TGs, and 59 standardization projects. Its first formal deliverable, "Exploring the Metaverse: Opportunities and Challenges," was endorsed in July 2023, highlighting the group's swift progress. This report examines the Metaverse's development, ecosystem, and related challenges and opportunities~\cite{westphal2023networking,wook2023standardization,ITU}.


\paragraph{IEEE Metaverse Standards Committee} The IEEE Metaverse Standards Committee\footnote{https://sagroups.ieee.org/metaverse-sc/}, which was previously known as the IEEE Virtual Reality and Augmented Reality Standards Committee, underwent an official name change on 21 September 2022 and expanded its scope of activities. This committee is dedicated to the development and international promotion of standards, best practices, and guidelines for the Metaverse, VR, and AR, following open and internationally recognized procedures. It is comprised of two primary working groups: the IEEE Metaverse Working Group (CTS/MSC/MWG) and the IEEE Mobile Device Augmented Reality Working Group (ARMDWG). The CTS/MSC/MWG has produced and approved the foundational P2048 standards entitled Terminology, Definitions, and Taxonomy Documents for the Metaverse in February 2023~\cite{westphal2023networking,Akbobek2023MetaPlatform,IEEE-Metaverse}.


\paragraph{World Metaverse Council} The World Metaverse Council~\footnote{https://wmetac.com/}, established on 1 October 2022, is committed to promoting an open, transparent, interoperable, and decentralized Metaverse. Its focus is on the development of standards and guidelines that ensure data security, uphold privacy, and protect individual rights, with a particular emphasis on establishing safeguards for children. Moreover, the Council advocates for the advancement of the Metaverse by supporting shared, open-source protocols, infrastructure, and financial systems, all aimed at cultivating an inclusive and collaborative Metaverse ecosystem~\cite{Akbobek2023MetaPlatform,WMC}.


\paragraph{Open Metaverse Interoperability Group} The Open Metaverse Interoperability Group~\footnote{https://omigroup.org/}, founded in April 2021, aims to develop protocols to connect virtual worlds and Metaverse. While specific standardization scopes are yet to be finalized, the broad objectives include identity, social graphs, inventories, transactions, avatars, 3D content, and portable scripted objects/scenes~\cite{wook2023standardization}.

\paragraph{ISO/IEC MPEG Working Group} MPEG~\footnote{https://www.mpeg.org/} serves as the primary working group responsible for the development of the international standard MPEG-V. Although MPEG is not a standards body in itself, it operates as Working Group 11 (WG 11) of Subcommittee 29 (SC 29) of the Joint Technical Committee 1 (JTC 1) on Information Technology under the joint auspices of the International Organization for Standardization (ISO) and the International Electrotechnical Commission (IEC). Since its establishment in 1988, MPEG has been instrumental in the formulation of international standards for the coding, compression, and transmission of audio, video, and related data. These standards are globally recognized and have significantly impacted the evolution of the digital media industry. In 2011, MPEG released MPEG-V, a framework that standardizes the exchange of information between interactive virtual environments and the physical world, enhancing interoperability through norms for sensory information exchange, data coding, and interactive device control~\cite{mpeg}.

\paragraph{Alliance for OpenUSD (AOUSD)} The AOUSD~\footnote{https://aousd.org/} was founded on August 1, 2023, by Pixar Animation Studios, Adobe, Apple, Autodesk, and the Joint Development Foundation (JDF) under the Linux Foundation. The JDF oversees its operations and management. USD, developed by Pixar, is a cutting-edge 3D scene description technology that promotes interoperability among tools, data, and workflows. The alliance aims to enhance 3D content creation interoperability through the USD framework, standardizing workflows for large-scale 3D projects and setting benchmarks for interoperability within the Metaverse ecosystem. On December 13, 2023, twelve industry leaders, including Cesium, Chaos, Epic Games, Foundry, Hexagon, IKEA, Lowe's, Meta, OTOY, SideFX, Spatial, and Unity, joined AOUSD, underscoring its industry support~\cite{linuxfoundation}. To achieve its mission, AOUSD announced a two-year development roadmap to establish USD as the global standard for 3D scenes and environments, focusing on facilitating interoperability among diverse data types~\cite{aousd2023deepdive}.

\paragraph{Metaverse Standards Forum (MSF)} Established on June 21, 2022, MSF~\footnote{https://metaverse-standards.org/} is a collaborative platform that unites Standards Development Organizations (SDOs) and companies to advance interoperability standards for the Metaverse. Its mission is to promote an open and inclusive Metaverse by fostering cross-organizational cooperation. Instead of creating standards independently, the MSF coordinates the efforts of various SDOs in areas such as 3D graphics, AR/VR, content creation, and geospatial systems. Participation in the forum is open and free, emphasizing a practical, action-oriented approach. This includes hosting prototyping events, developing open-source tools, and providing unified terminology and implementation guides to expedite testing and standard implementation. Initially, the MSF had 35 founding members, including Meta, Microsoft, and Nvidia. By August 2022, membership had expanded to over 1,200 organizations, highlighting the industry's commitment to an open-standards-based Metaverse, essential for unlocking its full potential. The MSF addresses significant interoperability challenges, outlines priorities for standards, and accelerates Metaverse technology standards' development and adoption. This enhances communication, reduces redundant work, and facilitates knowledge sharing. The Forum operates ten Domain Groups and has initiated three Exploratory Groups, focusing on initiatives like 3D Interoperability, Digital Twins, Geospatial Systems, Ecosystem Navigation, foundational Technology Stack, Education, and Legal and Policy issues. The MSF epitomizes an open and cooperative platform, striving for an accessible, inclusive, and extensively interconnected Metaverse ecosystem~\cite{Akbobek2023MetaPlatform,wook2023standardization,Hemphill2023MSForum,MSF}.


\subsubsection{Industrial Actions}
Research on interoperability within the industrial sector is still emerging. Abilkaiyrkyzy has conducted some initial organization by platforms~\cite{Akbobek2023MetaPlatform}, but a comprehensive study is still needed. One of the earliest recorded efforts was a 2018 initiative by IBM and Second Life, which interconnected two virtual worlds, allowing avatar transfers from Second Life to an OpenSim-based Metaverse~\cite{schonfeld2008ibm}. This was achieved using the Open Grid Protocol, a framework for virtual world interoperability. However, as reported by Techcrunch, the imperative lies in ensuring interoperability not only between virtual worlds but also with the Web, as it evolves towards a three-dimensional interface~\cite{schonfeld2008ibm}.

Recently, Decentraland~\footnote{https://decentraland.org/} has announced plans to collaborate with other crypto-Metaverses, implementing the IPSME protocol~\cite{Kim2021IPSME} to enable seamless asset retention and migration across Metaverses~\cite{farooque2021mana}. This protocol was highlighted during the 2023 Decentraland Metaverse Fashion Week, which involved platforms like Spatial~\footnote{https://www.spatial.io/} and Over~\footnote{https://www.overthereality.ai/}~\cite{schulz2022metaverse}.

Meta~\footnote{https://about.meta.com/metaverse/} has outlined plans to integrate Horizon Worlds~\footnote{https://horizon.meta.com/} and Crayta~\footnote{https://create.crayta.com/}, enabling shared avatars across these environments to enhance cross-platform interoperability~\cite{lang2022meta}. This aims to address the technical challenges of avatar portability across different systems, marking Meta's initial steps towards multi-platform integration~\cite{lang2022meta}.

Somnium Space~\footnote{https://somniumspace.com/} has partnered with HighFidelity~\footnote{https://www.highfidelity.com/} and JanusVR~\footnote{https://janusvr.com/} to create an interconnected VR world network called OASIS, using the VRBA open standard to promote platform interoperability. They are also working with Teslasuit to enhance tactile VR experiences with advanced suits and gloves~\cite{somniumtimes2023integration}.

Game engines like Unity~\footnote{https://www.unity.com} and Unreal~\footnote{https://www.unrealengine.com} have evolved beyond their original roles as development tools, becoming essential for fostering 3D interaction and innovation across various sectors to enhancing interoperability. Chia provides an in-depth analysis of these engines, noting their applications in education and industry~\cite{Chia2022GameEngines}. Unity and Unreal offer standardized tools and frameworks, reducing redundant coding, enhancing hardware compatibility, and enabling interoperability across applications. Their influence in application development and standard-setting is significant, impacting user operability and governance~\cite{Chia2022GameEngines}. However, more systematic research is needed to fully understand their roles in Metaverse interoperability.

Meanwhile, the formation of MSF and AOUSD while their rapid membership expansion among the leading enterprises, reflects a substantial market demand for seamless interoperability within 3D workflows~\cite{aousd_core_initiatives,branscombe2023openusd}. 

In summary, while research on interoperability in the industrial sector is still nascent, momentum is building. Initiatives like the MSF and OpenUSD Alliance and the growing importance of Unity and Unreal engines highlight this progress. Achieving a fully interoperable Metaverse is complex and requires ongoing research and development.


\subsubsection{Government Role and Policy Impact}
Gasser underscores the crucial role of governments and regulators in promoting interoperability within the digital ecosystem through policy tools. Governments can support standard development, address interoperability challenges, improve market transparency and competition via legislation, and reduce information asymmetry by mandating information disclosure. Their influence in public procurement can also promote interoperable solutions, and antitrust measures can prevent firms from withholding critical interoperability information, ensuring a competitive market. Despite criticisms about costs and efficiency, governmental involvement is essential for advancing interoperability~\cite{gasser2015intero}.

Yang's research delves into Metaverse governance via technical standards, focusing on creation, security, and compatibility. It highlights governments' critical role in leading Metaverse standardization and calls for collaborative efforts among governments, SDOs, and industry players to ensure secure and interoperable Metaverse growth. Recommendations include developing a comprehensive technical standards strategy, enhancing high-level design collaboration, and tailoring policies to industry-specific needs to address security and compatibility issues. Proactive standardization by the U.S., South Korea, Japan, Brazil, and China reflects a global commitment to Metaverse standards. This study suggests three steps: (1) Develop a detailed Metaverse standardization strategy with milestones, emphasizing safety standards and requirement collection; (2) Collaborate on high-level design to create an inclusive standards framework and harmonize policies for technological advancement; (3) Adapt strategies across platforms and industries, ensuring flexible responses, equitable rule enforcement, and intensified research on bridging digital and physical worlds~\cite{Yang2023TstandardGN}.

Akilli~\cite{akilli2022Turkiye} highlights Metaverse initiatives by South Korea and Turkey, though their impact on interoperability needs careful consideration. In early 2022, Seoul allocated several billion won to create Metaverse platforms, launching "Metaverse Seoul" to virtualize the city's lifestyle and culture, marking a new era in digital urban development. Similarly, Turkish President Erdoğan introduced the "Turkoverse" program to enhance regional integration within the Turkic world via the Metaverse. Akilli discusses whether these initiatives will aid or hinder a unified global virtual universe. While these efforts by South Korea and Turkey are forward-thinking, their specific role in promoting interoperability remains unclear. Akilli warns that without a focus on interoperability, the risk of a fragmented Metaverse landscape increases~\cite{akilli2022Turkiye}. Thus, although these countries are leveraging the Metaverse for cultural and economic gains, it is the structured planning and global collaboration on interoperability standards that will ultimately ensure a cohesive and interconnected metaverse.


\section{Current Status and Future Agenda}

\subsection{The Core and Status of Metaverse Interoperability}
We have clarified the essence of interoperability in the Metaverse through systematic deconstruction and analysis.

\subsubsection{From an infrastructure support perspective} Encompassing networks and devices, interoperability should facilitate connectivity for a broad spectrum of electronic devices. Although our core literature review revealed a limited number of related studies, this could be attributed to our deliberate exclusion of hardware-and-network-centric research to maintain a manageable scope, which may also indicate that this area has not yet attracted widespread research interest. We plan to delve deeper into this topic in our subsequent research. Current literature indicates a lack of a unified software framework to tackle cross-device functionality and interaction issues~\cite{tumler2022multi}. Nevertheless, exploratory research is underway, with recent studies beginning to address these challenges~\cite{tumler2022multi,yunsik2023xave,mike2023bgpblock}. Cedric Westphal's work also underscores active efforts toward networking standardization among network SDOs for the Metaverse~\cite{westphal2023networking}, signifying considerable potential for further inquiry in this field.


\subsubsection{From virtual worlds and Metaverse platforms standpoint} The development has progressed from initially connecting virtual worlds through middleware frameworks~\cite{Sergiy2013OpenMiddleware,Eric2014VWFramwork} like OpenSimulator~\cite{Cruz2015EVWIdenFeder} to developing and implementing advanced protocols and architectures like IPSME~\cite{Kim2021IPSME,farooque2021mana}. Metaverse platforms are constructed with both centralized and decentralized models, which beget distinct interoperability challenges: within centralized systems, among decentralized systems, and between the two models. Decentralization, often employing blockchain technology, further fragments interoperability into on-chain, off-chain, and hybrid on-chain/off-chain scenarios. The use of a single blockchain platform or multiple platforms amplifies these challenges~\cite{Yfang2023NetWeb3}. Although blockchain research has recently seen an upsurge, with studies examining cross-blockchain interoperability, including identity verification~\cite{siem2022selfsovereign,ying2023DIDs}, identity security~\cite{Sonali2023BCAuthenti,Ying2022AuthKeyAgree}, asset security~\cite{Jihyeon2023ASCTrading}, transaction efficiency~\cite{jiang2023fast,Yongjun2023HcnctBC}, and data management and protection~\cite{huynh2023blockchain}, the overall technological development for interoperability across platforms remains in the early stages within both industry and academia. 
There is a requirement for extensive and continuing research, not only to chart the overarching pathways among different interoperability categories but also to refine the exploratory studies currently underway. Our research acknowledges that full Metaverse interoperability among platforms is a long-term endeavor, but it has garnered significant attention, particularly in blockchain research. Despite the progress made, cross-platform interoperability is still in its infancy, necessitating substantial and dedicated research and development efforts to refine existing methodologies and navigate the complex web of interoperable connections among the rapidly proliferating Metaverse platforms.


\subsubsection{From the digital-physical fusion standpoint} Interoperability research highlights the potential of the Metaverse to facilitate more natural interactions and deeper integration with the real world than current Internet and IoT ecosystems allow. This field concentrates on the real-time mapping of real-world data to virtual entities and vice versa, facilitating dynamic interactions. Key technologies include sensors, IoT devices, digital twins, edge computing, federated learning, and semantic communication. Sensors and IoT devices form an information backbone extending to head-mounted displays (VR, AR, MR) and innovative wearables (tactile devices, brain-computer interfaces), integrated into physical environments for enhanced situational awareness. The merging of IoT and the Metaverse offers extensive technological and application opportunities~\cite{Jie2022EXRI,Jie2022ERIoT}. Digital twins, as precise digital replicas of physical entities, are poised to become central nodes for data processing and value generation~\cite{Yue2023ADynamicF}, though they face challenges like standardization, synchronization, integration, and security. Emerging solutions incorporate edge computing to reduce synchronization delays through localized data processing~\cite{Omar2023SynDTSubM,Lim2023MetaEdgeI}, federated learning to enhance data privacy and security via distributed processing~\cite{ali2023federatedAI}, and semantic communication to decrease bandwidth consumption by focusing on semantically processed data~\cite{Saeed2023SemCom}. The convergence of these technologies is crucial for enabling real-time interactions between virtual and physical worlds, thereby enhancing Metaverse interoperability. Future research aims to further harness this technological synergy, seeking novel solutions to issues like synchronization delays and data security, thereby driving continuous enhancements in the Metaverse ecosystem.


\subsubsection{From the data interoperability standpoint} Standardization is the key methodology to promote information exchange and system interoperability. Our study thoroughly reviews the evolution and discourse on data formats within the literature. 
VRML was an initial standard for 3D graphics, succeeded by X3D for varied encodings and web integration. glTF 2.0, known as the "JPEG for 3D," focuses on efficient 3D content transmission. USD handles complex 3D scenes with non-destructive workflows, and COLLADA promotes software-agnostic data exchange. These formats are compared in Table~\ref{tab:3d_format_comparison}.
MPEG-V is a widely-adopted standard fostering interoperability of sensory experiences across devices and virtual environments, facilitating immersive interactions. However, recent literature often overlooks current developments and challenges in data standard protocols. For example, interoperability among data formats remains critical. OpenUSD and Khronos Group are collaborating to address interoperability between USD and glTF 2.0~\cite{branscombe2023openusd}, but academic engagement is sparse.
Our study highlights a significant gap in the literature regarding comprehensive analysis and practical assessment of data format standards and their interoperability. We document the advancement of standardization, focusing on the leading SDOs in Metaverse interoperability, as detailed in Table~\ref{tab:SDOs}. Existing studies recognize their efforts, but a more systematic investigation is needed to fully understand their impact.
While standardization is crucial for interoperability, it also has drawbacks such as stifling swift innovation due to its slow-moving nature~\cite{Kim2021IPSME}. Alternative approaches to data interoperability are being explored, such as semantic communication for data extraction~\cite{Saeed2023SemCom}, knowledge graphs for data integration~\cite{Yfang2023NetWeb3,Rin2022FAIRKG}, and property graph patterns for new data modeling~\cite{Georgios2023Donna}. These alternatives are still nascent in the context of Metaverse interoperability.
High-level ontological views of hardware, software, and device components in the Metaverse exist, but fine-grained data models and interactions need further research. As Metaverse applications expand, clear data representation and models are essential for ensuring interoperability and scalability between the physical world, Metaverse, participants, devices, and events. Future research should focus on refining these aspects to enhance interoperability within the Metaverse.

\subsubsection{From industry practices and governance standpoint} Our study offers only a preliminary overview. Rapid industry innovation, such as the STYLE protocol~\cite{styleprotocol}, is outpacing academic research. Launched in 2022, the STYLE protocol facilitates the interoperability and monetization of virtual assets across the Metaverse, enabling seamless asset transfers via an NFT sub-licensing mechanism. Our review touches on initiatives by leaders like Second Life, Microsoft, Meta, Apple, and tools like Unity and Unreal, but it is not exhaustive. We plan to conduct a more detailed examination of industry movements for a comprehensive analysis in future research.
Government policies and actions are also superficially covered, with few publications addressing national strategies~\cite{Yang2023TstandardGN,akilli2022Turkiye}. Notable moves include Finland's "Metaverse Ecosystem Strategy" aiming for leadership by 2035~\cite{euractiv2023}, and China's initiation of a Metaverse standardization group in January 2024~\cite{CHGov2024SDG}. These developments highlight the need for continuous scholarly attention. Current policy discussions are fragmented and require more comprehensive exploration. Yang~\cite{Yang2023TstandardGN} suggests broad frameworks for government-industry collaboration but lacks explicit execution strategies. Future research should focus on detailed, action-oriented policy recommendations, balancing innovation with a secure, equitable, and sustainable Metaverse environment, particularly regarding interoperability challenges.



\subsection{Future Research Agenda on Metaverse Interoperablity}
Given the foundational understanding of Metaverse interoperability established in the first part of our findings, we now turn our attention to the future research agenda that can address the complexities and opportunities within this burgeoning field. Key areas of investigation include:

\subsubsection{Device and Platform Agnosticism}
Future research in Metaverse interoperability should prioritize robust frameworks for seamless interaction across a wide range of devices, including 2D platforms, VR and AR, holographic displays, and BCIs. The main goal is to develop adaptive software systems that intelligently conform to the diverse functionalities and performance constraints of different hardware.
Research should start with a systematic analysis of device characteristics, considering computational power, display technology, input methods, and sensory feedback. Establishing universal interoperability protocols will be crucial to guide software development, ensuring that applications can operate across various devices while optimizing performance to enhance each device's unique features. This will help deliver a consistent and engaging user experience.
Additionally, creating user-centric interfaces that are intuitive and adaptable to user preferences and contexts is essential. Software should detect if a user is interacting via a smartphone, VR headset, or BCI and adjust its interface and complexity accordingly. This adaptability is key for maintaining engagement and immersion in the Metaverse.
To balance innovation and usability, future research should develop standards and best practices for designing adaptive interfaces. These standards should be informed by extensive user testing and feedback to meet the needs of a diverse user base.


\subsubsection{Interoperable Virtual Environments} 
Seamless navigation across multiple sub-metaverses, each crafted by different developers, is a critical challenge for the Metaverse. Future research should focus on designing architectural standards and communication protocols to enable such fluidity. The goal is to enhance user experience and ensure continuity across diverse software platforms, whether centralized, decentralized, or hybrid.
Researchers must first understand the existing infrastructure of various sub-metaverses, mapping commonalities and differences. From this foundation, developing architectural standards for compatibility and interoperability is essential. These standards should address data formats, authentication methods, and asset transfer protocols to allow users to move between sub-metaverses effortlessly.
Protocols must be robust and flexible, supporting seamless transitions of user identities, digital assets, and social interactions. They should ensure that user actions in one sub-metaverse are reflected across others, maintaining user experience continuity. Additionally, a user-centric approach is crucial, involving intuitive navigation systems and user interfaces that are easy to operate. Privacy and security must also be prioritized to protect user data during transitions.
Research should also explore governance models, from corporate to community-led approaches, to understand their impact on user experience and the Metaverse's architecture. Collaboration among industry stakeholders, including technology developers, regulatory bodies, SDOs, and user representatives, is necessary. This collaboration will drive the creation of open standards that evolve with the Metaverse's growth.
In conclusion, future research should aim to establish an interoperable framework that supports smooth transitions between sub-metaverses. This requires a blend of technical standards, user experience design, governance considerations, and collaborative efforts to foster a cohesive and continuous digital ecosystem.


\subsubsection{Identity Management and Governance} 
The concept of 'identity' in the Metaverse is becoming crucial, requiring scholarly focus on its extension and management within phygital environments. Research should cover regulatory and normative elements, reflecting the blend of digital and physical realities. A comprehensive approach is needed to explore identity technologies, privacy concerns, and their impact on user behavior.
First, researchers must clearly define identity in the Metaverse, acknowledging its multifaceted nature as it transitions between real and virtual spaces. This includes legal recognition, portability across platforms, and universal verification standards to prevent fraud and ensure trust.
Second, the technological foundations of Metaverse identity need thorough investigation. This involves examining the potential of blockchain and other decentralized technologies for creating secure and immutable identity records, as well as exploring the role of artificial intelligence and other cutting-edge technologies in automating identity verification processes without compromising user privacy.
In terms of privacy, frameworks empowering users to control their personal information are essential. Research should evaluate current data protection regulations in the Metaverse context and propose necessary amendments or new policies. This includes examining self-sovereign identity models where users own and control their data.
The influence of identity management on user behavior must also be scrutinized. Users' perceptions of their identity in the Metaverse can affect interactions, social behaviors, and engagement. Ethical considerations should be integrated into identity systems to prevent misuse and ensure a respectful, inclusive environment.
A multi-stakeholder approach is vital, involving legal experts, technologists, sociologists, and users. This collaboration can produce comprehensive guidelines and best practices for identity management in the Metaverse ecosystem.
In summary, addressing identity complexities in the Metaverse is an interdisciplinary challenge. Future research should aim to construct a foundational framework for identity, ensuring a secure, private, and user-centric experience in the digital-physical convergence.


\subsubsection{Blockchain Integration and Interoperability} 
The ascent of decentralized platforms within the Metaverse has rendered the interoperability of blockchain technologies a subject of paramount importance. While blockchain technology has been extensively studied, there is a discernible lack of detailed studies on the nuanced hybrid interoperability models encompassing on-chain(intra-chain or inter-chain), off-chain, and hybrid models. Future research must address this gap by focusing on the integration and functionality of these models within the Metaverse ecosystem.
Research should begin with a comprehensive taxonomy of blockchain interoperability types, clarifying their distinctions and connections. This foundation will support further exploration into the strengths and limitations of existing interoperability solutions, particularly in terms of scalability, security, and speed, which are vital for a seamless Metaverse experience.
Next, research should investigate building hybrid models that enable complex, real-time interactions in the Metaverse. This includes standardizing smart contracts across blockchains to ensure seamless asset and identity transfers. Additionally, understanding how off-chain computation and data storage can interact with blockchain networks is crucial for managing the Metaverse's data loads without compromising decentralization principles.
Future studies must also consider the governance and regulatory implications of hybrid interoperability models. As the Metaverse grows, balancing compliance with international laws and standards while maintaining decentralization will be challenging. Collaboration between academia, industry, and regulatory bodies is essential to advance this research agenda. A multidisciplinary dialogue can foster innovative solutions that enhance interoperability while respecting blockchain's decentralized nature.
In conclusion, the current state of interoperability in the Metaverse's decentralized platforms requires focused academic inquiry. Future research should dissect complex hybrid models, emphasizing scalability, security, and regulatory compliance to meet the diverse demands of the Metaverse.


\subsubsection{Synthesis of Digital and Physical Worlds} 
The integration of sensors, IoT, digital twins, and advanced technologies like edge computing, federated learning, and semantic communication is vital for advancing the Metaverse. Research should focus on standardization, real-time synchronization, system integration, and security to ensure a seamless, secure, and efficient Metaverse experience.
First, the role of sensors and IoT devices as data collectors needs scrutiny. These devices bridge the physical and virtual worlds by collecting real-time data for digital twins. Research should aim to standardize data formats and communication protocols to ensure smooth information flow between devices and the Metaverse.
Second, edge computing can reduce latency and improve synchronization by processing data closer to its source. Studies should optimize edge computing infrastructure for the Metaverse to efficiently distribute computational loads.
Federated learning, which trains models on decentralized devices, can enhance Metaverse intelligence without compromising privacy. Research should explore its incorporation to personalize user experiences while maintaining data confidentiality.
Semantic communication, focusing on the meaning of messages, promises more intuitive Metaverse interactions. Studies should evaluate models that enable clearer, context-aware exchanges between users and systems.
Real-time synchronization is crucial for coordinating complex interactions across devices and platforms. Research should propose robust, scalable, low-latency synchronization mechanisms to keep digital twins up-to-date with their physical counterparts.
System integration across diverse technologies requires attention. The Metaverse's convergence of multiple systems necessitates research into integration frameworks that support interoperability and modularity, allowing for plug-and-play integration of new technologies.
Security is a paramount concern due to the increased attack surface from numerous connected devices and complex integrations. Research should develop robust encryption methods, secure authentication protocols, and anomaly detection systems to protect the Metaverse from cyber threats.
In conclusion, future research should adopt a multidisciplinary approach to integrate sensors, IoT devices, digital twins, and emerging technologies within the Metaverse. Focusing on standardization, real-time synchronization, system integration, and security will create a robust, responsive, and secure Metaverse infrastructure.

\subsubsection{Data Interoperability and Standardization} 
The proliferation of virtual environments and the Metaverse has highlighted the importance of data format standards and their interoperability. However, a detailed analysis of these standards and their interoperation complexities is lacking in scholarly literature. Future research should address this gap by examining existing standards like VRML, X3D, glTF 2.0, and USD, focusing on overcoming data interoperability challenges.
Research should start with a comprehensive review of each format's specifications, evaluating their strengths, weaknesses, and ideal use contexts. VRML and X3D, as precursors in the field, offer lessons in legacy system compatibility, while glTF and USD have emerged as frontrunners in modern, efficient 3D content delivery and scene description, respectively. Comparing these standards on extensibility, efficiency, and suitability for various Metaverse interactions is crucial.
A key research focus should be the challenges of data format interoperability, such as fidelity loss during conversion, high computational costs, and maintaining interactive functionalities. The goal is to propose solutions for seamless data exchange, enabling assets created in one format to be used in another without compromising quality or functionality. Potential solutions include developing universal translators or middleware and using advanced AI algorithms to automate and optimize conversions. Establishing best practices for content creators could also mitigate interoperability issues at the source.
Addressing these complexities requires collaborations beyond traditional academic research. Partnerships between industry standard development organizations and academic researchers are essential. These alliances can share knowledge, resources, and insights, leading to robust and applicable research outcomes. Workshops, symposia, and collaborative projects can build consensus on key issues and focus research efforts on impactful areas.
In essence, improving data format interoperability in the Metaverse demands a combined theoretical and practical approach, supported by strong academia-industry collaborations. This will ensure scientifically rigorous and industry-relevant outcomes, contributing to a more cohesive and interoperable Metaverse.

\subsubsection{Semantic Data Representation and Knowledge Graphs} 
The expansion of the Metaverse into diverse domains necessitates precise semantic data representation and modeling for interoperability and scalability. Knowledge graph methodologies offer a promising avenue for future research.
Research should focus on innovative methodologies in knowledge representation, constructing intelligent data models that capture complex relationships within the Metaverse. Such models need to define attributes and interactions that faithfully represent the diverse elements of the Metaverse—including users, devices, and events—and their counterparts in the physical world. These models must dynamically adapt to the evolving Metaverse, integrating new interactions and entities, and supporting semantic richness to enable machines to understand context and meaning.
Research must address how different data models, standards, and protocols can be harmonized to allow for seamless communication between various platforms and systems. Strategies for data conversion, alignment, and fusion should maintain data integrity and semantics.
Scalability is another key issue. As the Metaverse grows, data models must handle increased loads. Solutions might include intelligent knowledge graphs, distributed architectures, and efficient indexing and query-processing mechanisms. To ensure robust interoperability and scalability, it is also essential to investigate the role of machine learning and artificial intelligence in enhancing knowledge graphs. AI and machine learning can enhance knowledge graphs by automating relationship discovery, predicting user behavior, and personalizing experiences. AI can also bolster data security and privacy within these complex systems. The successful implementation of these research objectives will require a multidisciplinary approach. Collaboration among computer scientists, data engineers, cognitive scientists, and domain experts will ensure methodologies and models are technically sound and user-aligned. 
In summary, developing sophisticated semantic data models is critical as the Metaverse evolves. Future research should create intelligent, dynamic, and scalable knowledge graphs to integrate the physical and virtual worlds, supporting complex interactions and providing rich, seamless experiences.


\subsubsection{Industry Practices and Development Trends} 
The Metaverse's rapid advancement, propelled by industry innovation, has led to the development of protocols such as STYLE, which are pioneering yet remain on the periphery of academic scrutiny. The academic community shall play its pivotal role in conducting rigorous, systematic research to validate, critique, and enhance these industry-led initiatives. Firstly, scholarly attention should be directed towards thoroughly examining exploratory protocols like STYLE. This entails not only understanding their technical underpinnings but also critically assessing their efficacy, scalability, and security implications within the broader context of the Metaverse's infrastructure. The goal is to identify gaps and potential improvements that can be addressed through academic research, contributing to the refinement and robustness of these protocols. In parallel, research should be dedicated to systematic tracking and analysis of actions taken by industry companies. This involves creating frameworks for monitoring their development and implementation of Metaverse-related technologies, strategies, and standards. Such frameworks could be based on a set of criteria that includes technological innovation, market impact, user adoption, and regulatory compliance. Through systematic tracking, researchers can discern patterns and trajectories in industry behavior, which is critical for anticipating future challenges and trends. Furthermore, future research should extend beyond observation to actively anticipate challenges that may arise as the Metaverse grows. This proactive approach requires a forward-thinking mindset and predictive models to forecast potential technical and societal impacts. Researchers should consider the implications of emerging technologies, the scalability of new protocols, and the integration of diverse systems within the Metaverse.

\subsubsection{Policy Recommendations for Collaborative Development} 
Amidst the expansion of the Metaverse, the centrality of interoperability emerges as a foundational pillar upon which the efficacy and sustainability of this virtual ecosystem rests. The creation of comprehensive policies that guide the collaborative development of interoperability standards is vital. Such policies must be informed by targeted research and crafted to stimulate cooperation between governments, SDOs, and corporations. Policy recommendations must first delineate clear objectives for interoperability within the Metaverse, encompassing both technical and ethical dimensions. These objectives should foster an environment where diverse systems and platforms can seamlessly interact while adhering to a common set of principles that safeguard user interests and promote an open digital economy. Researchers in public policy and technological governance should conduct an in-depth analysis to identify potential points of convergence for different stakeholders. This analysis should result in high-level principles and actionable insights that inform policy frameworks promoting communication and collaboration. Security forms a cornerstone of these policy frameworks. As the Metaverse evolves, its interconnected nature becomes an attractive target for malicious actors. Policies must mandate the implementation of advanced cybersecurity measures, data protection standards, and privacy-preserving technologies that are robust yet flexible enough to adapt to emerging threats. Fairness in policy development is essential to avoid the monopolization of the Metaverse by a few dominant entities. Policies should aim to level the playing field, providing equal opportunities for smaller corporations and startups to innovate within the Metaverse. This includes ensuring transparent practices in data usage and fostering a competitive ecosystem that encourages diversity in content and services. At the same time, innovation and competition within the Metaverse must not be stifled by overregulation. Thus, policies should be crafted to stimulate creativity and economic growth. This can be achieved by ensuring that interoperability standards do not impose unnecessarily restrictive technical requirements and providing a regulatory environment that encourages entrepreneurial ventures. In conclusion, the development of interoperability within the Metaverse demands a coalition of efforts supported by astute policy-making. Researchers interested in the fields must offer clear, actionable insights that guide the creation of policies ensuring robust, fair, and sustainable interoperability. These policies must be dynamic, encompassing a balance between the imperatives of security and fairness, and the drive for innovation and competition. Such collaborative and forward-thinking policy development will be instrumental in shaping a Metaverse that is not only technologically interconnected but also equitable and resilient in the face of future challenges.

By addressing these research directions, scholars can contribute significantly to the understanding and advancing Metaverse interoperability, bridging the gap between theoretical exploration and practical application.


\section{Conclusion}
In this study, we explore the evolving concept of the Metaverse and its critical interoperability, a concept that currently lacks a universally accepted definition. Our work aims to address this gap by providing a comprehensive, systematic review of the literature on Metaverse interoperability. Utilizing Urs Gasser's interoperability framework for digital ecosystems, we have structured our examination of this multifaceted issue, thereby addressing our first research question (RQ1). Through a critical analysis of the literature, we identified three essential layers of Metaverse interoperability: (1) Compatibility Among Multiple Devices, (2) Seamless Navigation and Interoperability Among Platforms, and (3) Integrated Interaction Between Physical and Virtual Worlds. Our detailed analysis within these layers advances a nuanced understanding of the Metaverse, contributing fresh perspectives and delineating a clear pathway to address our second and third research questions (RQ2 and RQ3). Our comprehensive examination of the Metaverse includes primary academic concerns such as user experience needs, technological infrastructure, common data protocols, key SDOs, and recent industry developments. We have identified current challenges and outlined a future research agenda, building a foundation for ongoing academic inquiry and technological advancement in Metaverse interoperability, in response to our fourth research question (RQ4). Since the Metaverse remains at a nascent stage, interoperability is essential for future growth. We encourage scholars to engage with this field and pursue the outlined research avenues. Collective intelligence and sustained innovation are essential for the Metaverse to realize its potential as an interoperable digital ecosystem. We anticipate that our contributions will motivate further research that addresses technical barriers and involves broader societal, governance, and global digital economic impacts. In summary, the journey towards a fully interoperable Metaverse is complex and remains to be comprehensively charted. Nonetheless, through analytical framing, consensus theme identification, and systematic research integration, we expect to advance the discourse and development at the forefront of this digital revolution. 


\bibliographystyle{ieeetr}
\bibliography{bibtex/bib/sample-base}

\end{document}